\documentclass{aa}
\usepackage{natbib}           
\usepackage{graphicx}         
\usepackage{amssymb}          
\usepackage{ulem}         
\usepackage[usenames,dvipsnames,svgnames]{xcolor}
\usepackage{txfonts}

\usepackage{soul}
\newlength{\imsize}

\begin{document}

\title{Flux rope, hyperbolic flux tube, and late EUV phases in a non-eruptive circular-ribbon flare}

\author{S.Masson\inst{1} \and E. Pariat\inst{1} \and G. Valori\inst{2} \and N. Deng\inst{3,4} \and C. Liu\inst{3,4} \and H. Wang\inst{3,4} \and H. Reid\inst{5}}

 \institute{LESIA, Observatoire de Paris, PSL Research University, CNRS, Sorbonne Universit\'es, UPMC Univ. Paris 06, Univ. Paris Diderot, Sorbonne Paris Cit\'e, \email{sophie.masson@obspm.fr} 
 \and University College London, Mullard Space Science Laboratory, Holmbury St. Mary, Dorking, Surrey, RH5
6NT, U.K. \and Space Weather Research Laboratory, New Jersey Institute of Technology University Heights, Newark, NJ 07102-1982, USA \and Big Bear Solar Observatory, New Jersey Institute of Technology, 40386
North Shore Lane, Big Bear City, CA 92314-9672, USA
\and SUPA School of Physics and Astronomy, University of Glasgow, G12 8QQ, UK}

 \abstract
   {The dynamics of ultraviolet (UV) emissions during solar flare provides constraints on the physical mechanisms involved in the trigger and the evolution of flares. In particular it provides some information on the location of the reconnection sites and the associated magnetic fluxes. In this respect, confined flares are far less understood than eruptive flares generating coronal mass ejections.}
   {We present a detailed study of a confined circular flare dynamics associated with 3 UV late phases in order to understand more precisely which topological elements are present and how they constrain the dynamics of the flare.}
   {We perform a non-linear force free field extrapolation of the confined flare observed with the HMI and AIA instruments onboard SDO. From the 3D magnetic field we compute the squashing factor and we analyse its distribution. Conjointly, we analyse the AIA EUV light curves and images in order to identify the post-flare loops, their temporal and thermal evolution. By combining both analysis we are able to propose a detailed scenario that explains the dynamics of the flare.}
  {Our topological analysis shows that in addition to a null-point topology with the fan separatrix, the spine lines and its surrounding Quasi-Separatix Layers halo (typical for a circular flare), a flux rope and its hyperbolic flux tube (HFT) are enclosed below the null. By comparing the magnetic field topology and the EUV post-flare loops we obtain an almost perfect match 1) between the footpoints of the separatrices and the EUV 1600~\AA{} ribbons and 2) between the HFT's field line footpoints and bright spots observed inside the circular ribbons. We showed, for the first time in a confined flare, that magnetic reconnection occured initially at the HFT, below the flux rope.
Reconnection at the null point between the flux rope and the overlying field is only initiated in a second phase. In addition, we showed that the EUV late phase observed after the main flare episode are caused by the cooling loops of different length which have all reconnected at the null point during the impulsive phase.}
   {Our analysis shows in one example that flux ropes are present in null-point topology not only for eruptive and jets events, but also for confined flares. This allows us to conjecture on the analogies between conditions that govern the generation of either jets, confined or eruptive flares.
}


\titlerunning{Non-eruptive circular ribbon flare}
\maketitle

 \section{Introduction}
 \label{Introduction}

Circular-ribbon flares received an enhanced interest over the last ten years \citep{Ugarte-Urra_al07,Masson_al09b,Reid_al12,Wang_al12,Dai_al13,Deng_al13,Jiang_al13,Jiang_al14,Jiang_al15,Liu_al13,Liu_al15,Sun_al13,VemareddyWiegelmann14,Wang_al14,Joshi_al15,Kumar_al15,Yang_al15,Janvier_al16}. They belong to a sub-class of flares for which one of the ribbon has an almost-fully closed quasi-circular or quasi-ellipsoidal shape. The magnetic field distribution of active regions where circular-ribbon flares are observed consists, in the simplest case, of a parasitic polarity  embedded in an opposite sign polarity of a larger dipolar active region. This circular ribbon is usually located slightly outside - closely surrounding - the parasitic polarity.  Some circular-ribbon flare regions can however present a more complex distribution \citep{VemareddyWiegelmann14,Joshi_al15}. In addition to the ellipsoidal-shaped ribbon, circular-ribbon flares are often associated with two other elongated ribbons. The inner ribbon is located within the circular ribbon in the parasitic polarity, while the second elongated ribbon, the outer ribbon, is located outside of the circular ribbon in the external polarity region, often at a relatively remote location.

Topological studies allow us to analyse the properties of the magnetic field and their relation with the flare ribbon geometry and dynamics \citep[e.g.][]{GorbachevSomov88,Mandrini_al91,Mandrini_al95,Mandrini_al97,Demoulin_al94b,Demoulin_al97,Savcheva_al12,Savcheva_al15,Savcheva_al16}. Circular-ribbon flares are tightly linked with the topological structure of a 3D null-point \citep{Masson_al09b}. A null point defines a dome-like shape separatrix surface, the fan, and two singular field lines, the spines, originating from the null-point  each of them belonging to one of the semi-space bounded by the fan surface \cite[cf][]{Longcope05,Pariat_al09}. In a C-class circular ribbon flare \citet{Masson_al09b} showed that the circular ribbon corresponds to the footpoints of the fan separatrix surface, whereas the inner and outer ribbons respectively correspond to the spine footpoint within and outside of the closed fan surface. \citet{Masson_al09b} also highlighted that the elongated shape of the inner and outer ribbons is caused by reconnection across the Quasi-Separatrix Layer (QSL) which surrounds the null point separatrices. 
Such a structure is referred to as a QSL halo around the separatrices \citep{Masson_al09b,Pontin_al16}.  \cite{Pontin_al13,Pontin_al16} showed that for null point topology magnetic reconnection in the null-point QSL-halo occurs in the current sheet surrounding the null point. In this sense, both null-point and QSL-halo reconnection are the same. However, null-point and QSL-halo reconnection lead to a different change of connectivity which may be significant to interpret observational signature and the dynamics of the flare. In term of connectivity changes, null-point reconnection leads to a jump of connectivity between the inner and the outer fan connectivity domain while QSL-halo reconnection implies continuous changes  within the same connectivity domain.

Solar flares are divided into two classes : the eruptive and non-eruptive/compact or confined events. During an eruptive flare, a magnetic structure is ejected and is later observed as a coronal mass ejection (CME) in white-light coronagraph images. On the contrary, non-eruptive (or compact) flares do not generate any CMEs\footnote{Flare inducing jets are here considered as non-eruptive flare.}. Eruptive flares are associated with a magnetic flux rope formed either before or during the flare. This magnetic structure plays a central role in the ejection dynamics and is also at the origin of the Torus instability trigger scenario \citep{TorokKliem07}. 
Searching for the existence of magnetic flux rope before the onset of eruptive flares is therefore one of the main driver of studies investigating the solar eruption trigger mechanisms. Several circular-ribbon flares present an eruptive nature  associated with a flux rope.\citep{Dai_al13,Sun_al13,Jiang_al14,
Liu_al15,Yang_al15,Janvier_al16}. 
 In addition to the flux rope, those circular-ribbon eruptive flares have a null-point topology which is essential for the breakout trigger scenario \citep{Antiochos_al99}. Therefore, eruptive circular-ribbon flares are a particularly interesting type of events since multiple topological structures involved in different eruption scenarios are present. 

Do non-eruptive/compact circular-ribbon flares present twisted flux rope enclosed below the fan surface? To this day no study has ever thoroughly investigated the magnetic topology of compact circular-ribbon flares. Therefore, compact flares are generally not considered as being related to magnetic reconnection involving sheared or twisted flux tubes. 


A sub-class of circular-ribbon flares are associated with jets \citep{Wang_al12,Liu_al13}. Solar jets are defined as a collimated and mainly radially propagating plasma which is observed either in emission or in absorption. Most of the jet models agree that they originate in 3D null-point topology \citep[see review by][]{Raouafi_al16}. It is more and more widely accepted that jets are triggered by magnetic reconnection between a flux rope, initially confined below the fan, and the outer-fan magnetic field \citep[e.g.][]{Pariat_al15a,Raouafi_al16}. Jets are indeed observed to be associated with micro-sigmoids \citep{Raouafi_al10} and erupting filaments \citep{Sterling_al15} suggesting the presence of a flux rope structure.
Since flux ropes are found in 3D null-point topology associated with non-eruptive flares accompanied by jets  and circular-ribbon eruptive flares, we could expect to find them at all scales in systems with a 3D null-point and thus in compact jet-less circular ribbon flares. The existence of flux ropes below a 3D null-point dome would be particularly important for the reconnection dynamics in the system.

Another signature of the complex dynamics of the reconnection in null point systems is the observed brightening propagation along the forming circular ribbon \citep{Masson_al09b,Reid_al12,Wang_al12,Sun_al13,Jiang_al15}. From the data-initiated and data-inspired driven 3D MHD simulation, \citet{Masson_al09b} demonstrated that this propagation is very-likely due to the slipping reconnection in the QSL-halo surrounding the null separatrices. \citet{Sun_al13}  observed this slippage by carefully looking at the dynamics of hot post-flare loops around the spine during a circular ribbon flare. Those bright loops are observed when the plasma density increases in the reconnected flux tubes. The density increase may be caused by the evaporation of chromospheric material along the flux tube \citep[eg.][]{DelZanna08} or rarefaction waves \citep{Bradshaw_al11}. Therefore, the spatial evolution of the post-flare loops can be used as an indicator of the evolution of the reconnected flux \citep{Aulanier_al07,Masson_al14,Malanushenko_al14,Dudik_al14}.

It is common to observe apparent slipping of post-flare loops. Such observation is attributed to 3D magnetic reconnection of neighbouring field lines belonging to quasi-separatrix layers \citep{PriestDemoulin95,Aulanier_al06}. Studies of magnetic reconnection at and around 3D singular null point is an active field of theoretical and numerical investigations to understand the transfer of the magnetic flux \citep{Pariat_al09,Pariat_al15a,HachamiPontin10,Masson_al12a,Fuentes_al12,Fuentes-FernandezParnell13,Pontin_al13,Pontin_al16,
WyperPontin14a,WyperPontin14b,WyperDeVore16}. In the case of eruptive flares, numerical simulations of the eruption of a flux rope under a null \citep{Lugaz_al11,Jiang_al13} show a complex evolution of the connectivity as the flux rope reconnects/interacts at the null. Combined observations of the ribbon dynamics and the 
flare-loop evolution in circular ribbon flares are essential in order to provide constraints for these theoretical works. 

Finally, flares involving a null point are particularly prone to EUV emissions in late phases \citep{Dai_al13,Sun_al13,Liu_al13,Li_al14}. EUV late phases have been first noticed by \citet{Woods_al11} and correspond to the existence of at least one distinct flux peak in the EUV irradiance time-profile after the impulsive phase of some flares, with no correspondence in X-ray \citep[see also][]{Hock_al12,Liu_al15}. Several explanations have been put forward to explain these late phases : additional heating in the late-phase loop  \citep{Woods_al11,Hock_al12,Dai_al13,Liu_al15}, the differential cooling of post-flare loops of different lengths \citep{Liu_al13,Li_al14} or a combination of the two \citep{Sun_al13}. The multi-wavelength analysis of new flaring regions presenting EUV late phase is therefore crucial to complete our understanding of the plasma thermodynamics during the gradual phase of flares.

Compact circular-ribbon flares are particularly interesting flares to study for their magnetic structures, their 3D properties of magnetic reconnection and the plasma response to it. In the present study, we build on the observed event studied by \citet{Deng_al13}. They analysed an unprecedented high-resolution and high-cadence $\textrm{H}\alpha$ imaging observations of a C4.1 circular-ribbon flare and showed the detail of the outer ribbon dynamics and confirmed its formation mechanism resulting from electron beam heating. We undertake here a complementary study of this event by performing a detailed examination of the magnetic topology of this flaring region with respect to the EUV emission both during the impulsive and the gradual phases.

 We first present the EUV solar ribbons and the EUV light curves evolution of the flare in which we identify a main phase and 3 EUV late phases (\S~\ref{confined flare}). We computed a NLFFF extrapolation and analyse the magnetic topology (\S~\ref{magnetic topology}). In section~\ref{detailed flare dynamics} and \ref{EUV late phase} we combined the topological analyses and the temporal and spatial evolution of the bright post-flare loops to interpret the dynamics of the compact flare, from the pre- to the post-flare phase. We summarise and discuss our results in  \S~\ref{conclusion}.

 \section{Confined flare on Oct 22, 2011}
 \label{confined flare}
 
 \subsection{UV ribbons}
 \label{UV ribbons}

 On October 22, 2011, the compact flare SOL2011-10-22-1521 was observed in AR ~11324 between 15:00~UT and 18:00~\rm{UT} with the AIA imager \citep{Lemen_al12} onboard SDO \citep{Pesnell_al12}. No CME is observed. The high temporal cadence (12 s) and the high spatial resolution ($1.2^{\prime \prime}$) of the AIA images allow us to study in details the evolution of the flare.

\begin{figure}
\setlength{\imsize}{0.49\textwidth}
   \includegraphics[width=\imsize,clip=true]{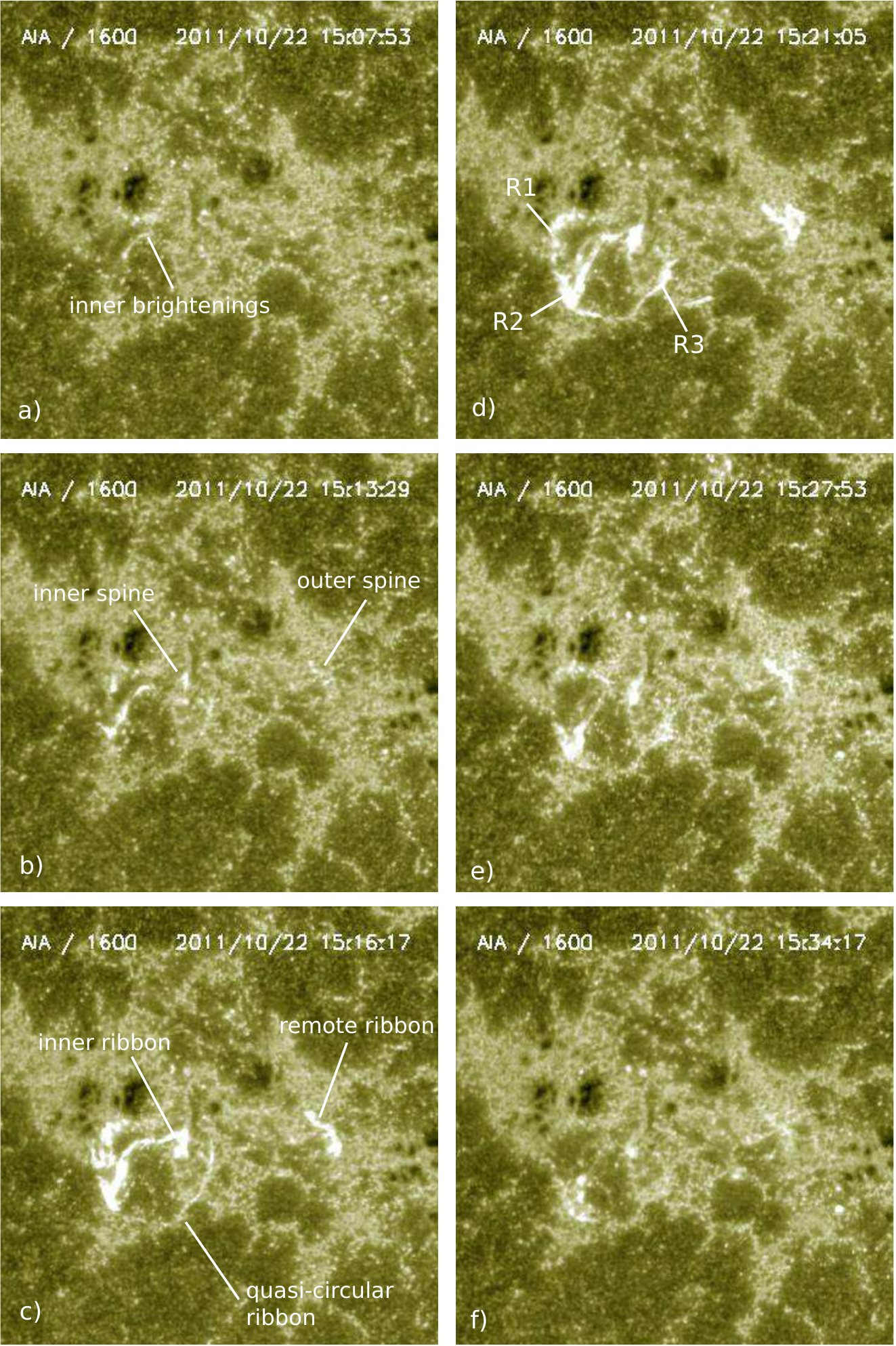}
\caption{Time sequence of AIA images at 1600~\AA{} observed in AR 11324 during the flare on October 22, 2011. The field of view is $x \simeq [-450, -250]^{\prime \prime}$ and $y \simeq [0, 200]^{\prime \prime}$. Panel a) shows brightenings observed prior to the null-point reconnection. On panels b) and c) are identified the 3 main ribbons related to null-point reconnection : quasi-circular , inner and outer ribbons associated with the fan, inner and outer spine separatrix footpoints. On panel d), R1, R2 and R3 denote the 3 brightest segments along the quasi-circular ribbon. Panels e) and f) show the decaying phase of the flare. The full temporal evolution is available in an online movie that also shows the AIA sequence in the 131~\AA{} and 171~\AA{} channels. (The movie has not been corrected from the solar rotation.)}
 \label{fig1}
\end{figure}

Figure~\ref{fig1} shows the temporal evolution of the flare at 1600~\AA. We identify a quasi-circular ribbon which encloses one inner ribbon and several bright spots; and a remote ribbon is located westward of the circular ribbon (see labels on Figure~\ref{fig1}, left column) at about $35~\rm{Mm}$ from the fan. Such ribbon morphology corresponds to a null-point flare \citep{Masson_al09b,Reid_al12,Wang_al12,Dai_al13,Liu_al13,Deng_al13,Sun_al13}.  Similarly to previous studies, we expect that the quasi-circular ribbon traces the location of the fan separatrix footpoints and its QSL-halo, that the inner and outer ribbons are respectively located at the inner and the outer spine singular field line footpoints, and, that the extension of these latter ribbons is caused by the QSL-halo around the spines.

The main phase of the flare occurs between 15:13~UT and 15:28~UT, when we observe the brightening of the 3 ribbons  (Figure~\ref{fig1}c \& movie 1 of the online material). At 15:13~UT, fragments of the circular ribbon brighten simultaneously and two bright kernels  appear inside and westward of the circular ribbon (see labels on Figure~\ref{fig1}). The simultaneous appearance of the inner and remote bright kernels suggests (and is confirmed by our topological analysis in \S~\ref{magnetic topology}) the anchoring of the inner and outer spines, and indicates that field lines enclosed below the fan have reconnected at the null point with the outer magnetic flux, i.e., the null-point reconnection has started. At 15:14~\rm{UT}, bright segments of the quasi-circular ribbon become more intense, starting in the north-east part. Then brightenings extend following the circular shape in a counterclockwise direction and finally the western part of the quasi-circular ribbon gets brighter too. Even though most of the quasi-circular ribbon brightens, we identify 3 main segments along it that are brighter than the rest:  R1 on the north-east segment,  R2 on the east and R3 lying along the west part (see labels on Figure~\ref{fig1}d). Meanwhile, the inner and the remote bright kernels become brighter and extend to form two elongated ribbons. After 15:16~UT, the remote ribbon shape changes and a "tail" develops westward of the initial elongated ribbon. At 15:21~\rm{UT}, when the flare reaches its maximum intensity, the remote ribbon has an inverse-Z shape \citep[see for details][]{Deng_al13}. 

The spatial and temporal evolution of the ribbons are fully consistent with the standard null-point/slipping reconnection model \citep{Masson_al09b}. However, before the main null-point episode, bright spots are observed at 1600~\AA{} inside the quasi-circular ribbon. These brightening are first observed at 15:07~UT (Figure~\ref{fig1}a) and remain the only visible brightenings until the onset of the null-point reconnection at 15:13~UT (Figure~\ref{fig1}b). Those internal bright spots do not, {\it a priori}, result from null-point reconnection even though they are located inside the quasi-circular ribbon. They are not predicted by the standard null-point/slipping reconnection model. What is the origin of those bright spots appearing inside the fan prior to any null-point related ribbon ?  Do they play a role in  triggering the null-point flare ?

 \subsection{EUV light curves}
 \label{thermal evolution}
 
\begin{figure}
\setlength{\imsize}{0.49\textwidth}
   \includegraphics[width=\imsize,clip=true]{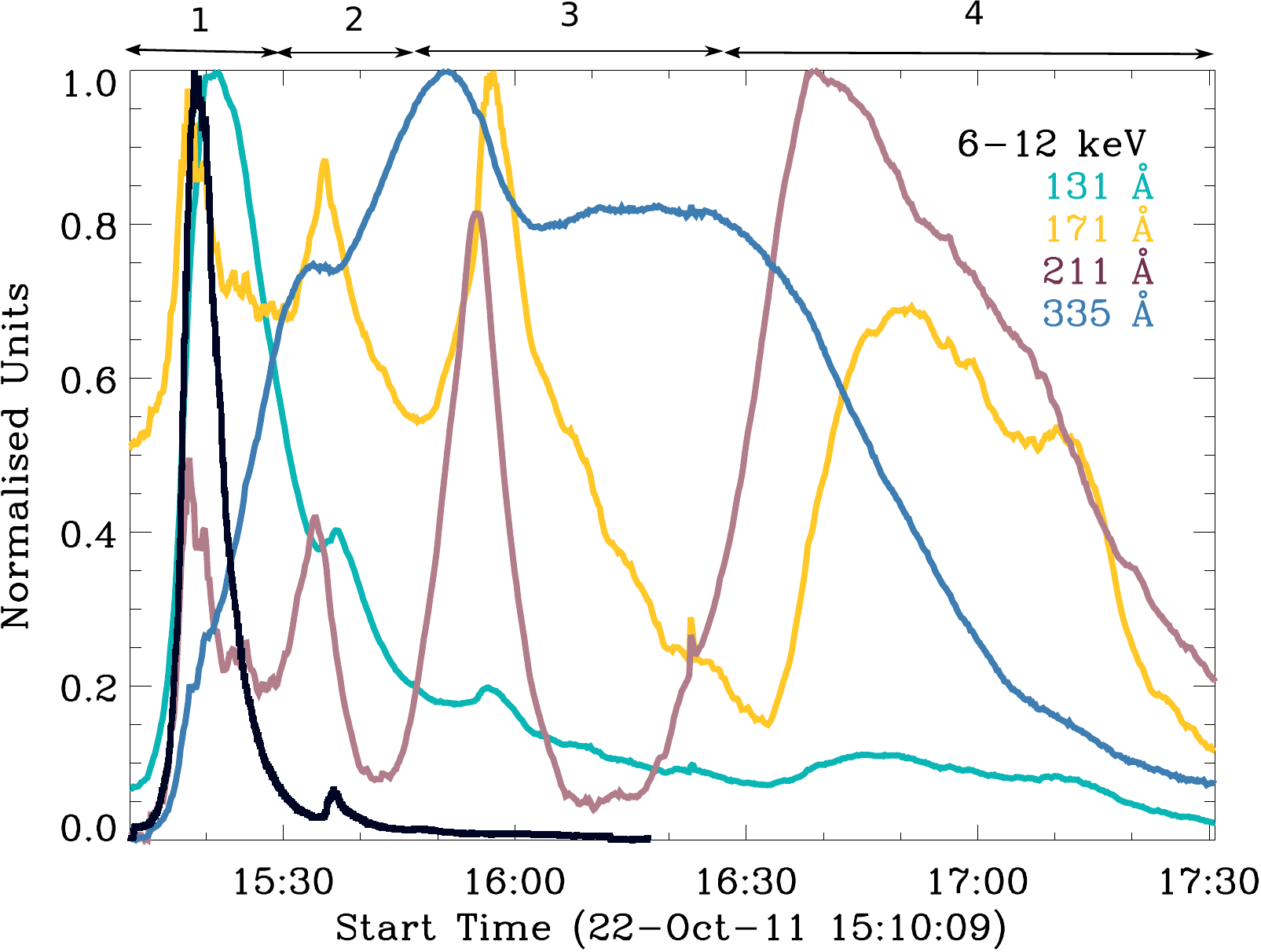}
\caption{Multi-wavelength light curves of the 8 AIA EUV lines and the X-ray light curve at $6-12~\rm{keV}$ from RHESSI  computed in a box located at the longitude $x=[-450,-250]^{\prime \prime}$, and the latitude $y = [0,200]^{\prime \prime}$. Their time profile is structured by a main flare episode between 15:10~UT and 15:28~UT, and three late episode peaking respectively around 15:34~UT, 15:54~UT and between16:39~UT and 16:50~UT}.
  \label{fig2}
\end{figure}

Using the AIA capacity, we determine the multi-wavelength evolution of the flare. Figure~\ref{fig2} shows the light curves of the flaring region ($x=[-450,-250]^{\prime \prime}$, $y = [0,200]^{\prime \prime}$) for 4 EUV coronal AIA channels and the soft X-ray light curve ($6-12~\rm{keV}$) observed by RHESSI \citep{Lin_al02}.
According to the response function of AIA, the observed lines can be divided in 3 groups : cold, warm and hot \citep[see][their Table 2]{Reale10}. The 335~\AA{} is a hot channel with a temperature response function peaking at $T=10^{6.4}~\rm{K}$.
The 131~\AA{} line has two hot components at $T=10^{7.0}~\rm{K}$ and $T=10^{7.2}~\rm{K}$ and a cold component at $T=10^{5.6}~\rm{K}$.  The 211~\AA{} lines corresponds to a warm emission and its temperature response function peaks  $T=10^{6.2}~\rm{K}$. The 171~\AA{} channel corresponds to a cold plasma at $T=10^{5.8}~\rm{K}$. Except the 335~\AA{} lines which has a smooth and long duration time profile from 15:07~UT to 18:00~UT, the other time profiles are structured by several well pronounced peaks.

We identify 4 major episodes between the 15:07~UT and 18:00~UT, defined by the presence of peaks of luminosity in at least 2 channels (see labels on Fig.~\ref{fig2}). During the main episode (15:07~UT - 15:24~UT), all the channels rise and most of them reach their maximum at 15:18~UT, except 335~\AA{} that shows a local maximum at 15:18~UT but reaches its absolute maximum later. During this main episode, we observe a peak in the soft X-ray emission ($6-12~\rm{keV}$).  This first episode is temporally consistent with the null-point reconnection episode identified at 1600~\AA{} (see \S~\ref{UV ribbons}). 

After the main episode, three sharp peaks are observed in the warm (211~\AA{}) and cold lines (171~\AA{}) and are accompanied by a local maximum in the hottest line (335 \AA{}). Those peaks appear around  15:34~UT, 15:54~UT and between 16:39~UT and 16:50~UT, respectively about 16, 36 and 81 minutes after the first peak. In each of those episodes, we observe the same sequential evolution of the peaking time in function of the line temperature. 
 The local maxima of the first EUV late phase peak  in function of the plasma temperature :  the 211~\AA{} emission at 15:34~UT, the 193~\AA{} emission at 15:35~UT(not shown here), and  the 171~\AA{} emission at 15:36~UT. For the second EUV late phase, the light curve at 211~\AA{} locally peaks at 15:54~UT,  the 193~\AA{} emission at 15:55~UT (not shown here), and the 171~\AA{} emission peaks at 15:56~UT (see Figures~\ref{fig2}). The time delay for the third EUV late phase is much more important. Indeed the emission at 211 \AA{} peaks locally at 16:39~UT, the 193~\AA{} emission at 15:43~UT (not shown here), and the emission at 171 \AA{} peaks at 16:50~UT .
The first post-flare EUV episode ( $\simeq $ 15:34~UT) is associated with chromospheric brightenings in the vicinity of the inner and remote ribbon as well as along the East part of the quasi-circular ribbon (R2). The second ( $\simeq$ 15:54~UT) and the third EUV peak (from 16:39~UT to 16:50~UT) are not associated with any well-defined chromospheric brightenings, but some sparkling spots can be seen until 15:53~UT (see left panel of the online animation).

EUV late phase has been first identified by \cite{Woods_al11} and has been defined by 4 criteria. Since then, several other  studies addressed the origin of late EUV phases \citep{Liu_al13,Sun_al13,Dai_al13} and the initial definition as been adjusted. According to those studies, the minimalist definition for a late EUV phase can be a warm (211~\AA{}, 193~\AA{}) and cold (171~\AA{}) EUV emission peaking several minutes to hours after the main flare phase, with no soft X-ray counterpart but associated with distinct emitting EUV loops. 
 
For the present flare, the AIA 335~\AA{} emission rises slower and later than other channels. It reaches a local maximum 2-4 minutes before each of the 3 peaks and its temporal structure corresponds to the EUV late phase described above. While it is common to use EVE FeXVI 335 line to identify a late EUV phase, our event is not strong enough at the flaring time to create a clear emission peak in the whole Sun EVE irradiance. However, all previous events showed that the late EUV phase is equally observed with EVE and AIA \citep{Woods_al11,Dai_al13,Sun_al13}.  For the present flare, no significant enhancement of RHESSI soft X-ray and hot emission are detected (Figure~\ref{fig2}), however the flare class and the X-ray flux are much lower than in previously reported late EUV phase flares. The additional X-ray flux of this region may be covered by the overall flux of the Sun. 


 \section{Magnetic Model of the Circular Flare}
\label{magnetic topology}

 So far, identifying the topological elements involved during solar flares, eruptive or not, has been a successful way to understand the origin and the evolution of the solar eruption. As we increase the complexity of the extrapolation model of the magnetic field, we obtain more and more details on the flare. In the following, we analyse the magnetic topology of the flare in order to identify the magnetic topological elements and determine their role in the flare dynamics. 

 \subsection{Non-Linear Force Free Extrapolation}
 \label{NLFFF extrapolation}
 
The HMI instrument \citep{Scherrer_al12} provides a highly resolved magnetogram (time cadence is $45~\rm{s}$ and spatial resolution of $1^{\prime \prime}$). However, the vector magnetogram requires a longer integration time, the cadence is 12 minutes, and $0.5^{\prime \prime}$ is the CCD pixel size. The active region consists in a East/West oriented bi-polar configuration with a negative parasitic polarity embedded in the eastern positive polarity (left panel, Figure~\ref{fig3}). 

 The vector magnetogram at 15:00~UT of corresponding to AR11324 is included in the patch 976 of the HARP catalogue \footnote{\url{http://jsoc.stanford.edu/doc/data/hmi/harp/harp_definitive/2011/10/22/harp.2011.10.22_15:00:00_TAI.png}}. The standard data product of the HARP pipeline as of 25 January 2013 was used. This includes the removal of the ambiguity in the direction of the transverse field, as well as Cylindrical Equal Area projection remapping \citep[see][for more details]{Hoeksema_al14}. The area of interest was extracted from the HARP patch data, the resolution was halved using a flux-conserving coarsening, and a median smoothing with 7-pixel boxcar was applied to all three field components. Since the magneto-frictional relaxation equations are parabolic in nature, the lower resolution allowed us to obtain a deeper relaxation level in an acceptable running time. The median-smoothing was applied in order to eliminate some of the salt-and-pepper fluctuations, present especially  in the area of low field/low signal-to-noise ratio.

Vector magnetograms are estimations of the magnetic field inferred from the inversion of spectropolarimetric profiles that are remotely measured. The origin of the emission for the SDO/HMI sprectropolarimeter is photospheric, \textit{i.e.} coming from a region of high-$\beta$ plasma \citep{Lites00}. Therefore, in order to use vector magnetograms as boundary conditions of NLFFF extrapolation codes, it is advantageous to modify the observations in such a way that Lorentz forces in the magnetogram are reduced (i.e., preprocessing, e.g.  \cite{Schrijver_al08}). We employ the technique of \cite{Furhmann_al07}, which allows us to fix a limit to the modification of each component separately. In particular, the preprocessing is applied to the horizontal components only, with a maximum allowed variation of the measured horizontal components corresponding in each pixel to the larger between  $100~\rm{G}$ and 50\% of the local value. These maximal ranges of variation result into an average modification of $52~\rm{G}$ (respectively, $61~\rm{G}$) in the $B_x$ (respectively, $B_y$) component,  and in a decrease of Lorentz forces on the magnetogram from $0.07$ before preprocessing to  $10^{-3}$ after preprocessing, according to  the definition used in \cite{Metcalf_al08}.

The preprocessed vector magnetogram is then extrapolated to build the coronal field model using the implementation of the magneto-frictional method described in \cite{Valori_al10} with open lateral and top boundaries and on 3 levels of successive grid refinement. The resulting numerical model has localised relatively high forces, possibly due to the fact that, at the time of the observation, the system is close to the dynamical phase. The field then is less force-free in volumes directly connected to non-force-free areas of the magnetogram. The resulting force-freeness level, as defined by \cite{Wheatland_al00} is $\sigma_J\equiv\int  |\vec{J}_\perp|  \, d\mathcal{V}/ \int  |\vec{J}|  d\mathcal{V} =0.44$.


One of the consequences of the boundary condition not being perfectly compatible with the force-free assumption is that the resulting extrapolated field presents local violations of the solenoidal property of the field. Such violations can be quantified as a fraction of the total energy, in this case equal to $E_{\rm ns}/E=0.07$  according to the notation of  \cite{Valori_al13}. The presence of a finite  field divergence is potentially a source of mismatch between field lines and observations, but, as the detailed comparison with observations below shows, its distribution in the volume appears to largely  average away.

 \subsection{Null-point topology}
 \label{Null point topology}

\begin{figure*}
 \setlength{\imsize}{0.99\textwidth}
   \includegraphics[width=\imsize,clip=true]{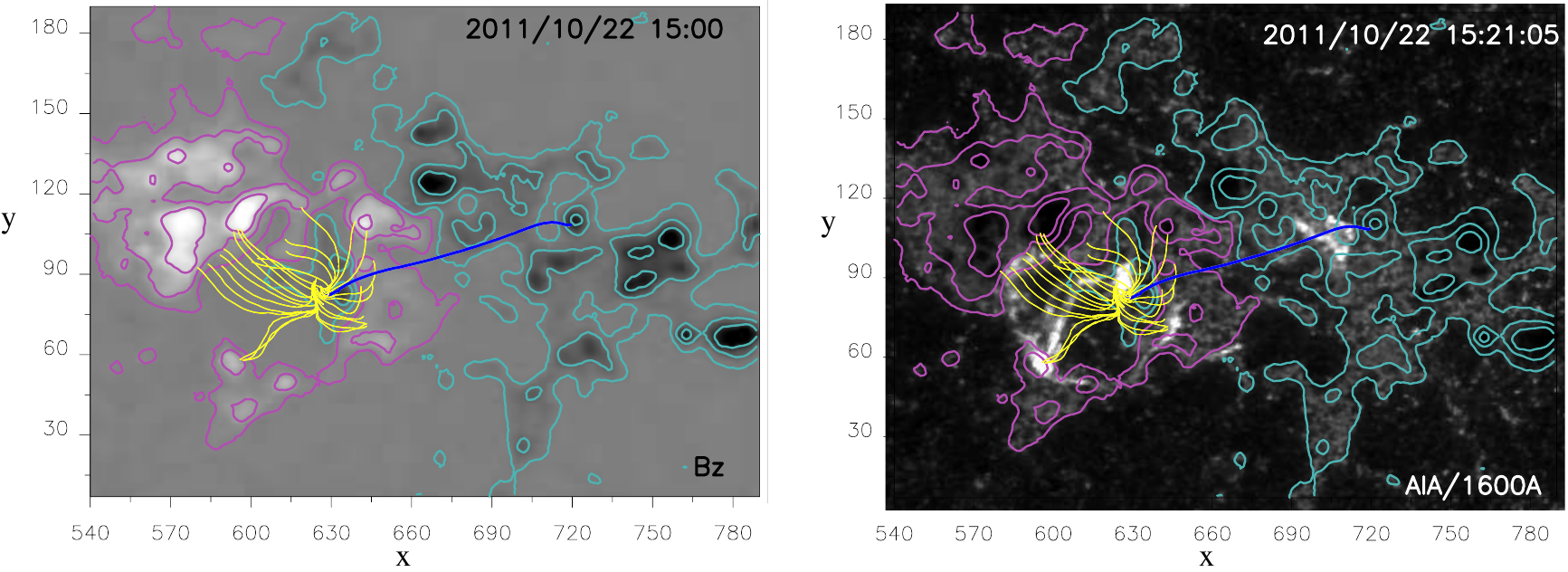}
\caption{ Null-point topology resulting from the NLFF extrapolation. Left panel: The line of sight magnetic field is grey-shaded. Pink and blue iso-contours correspond to positive and negative $B_{los} = \pm 40, 250, 750~\rm{G}$, respectively. Blue and yellow field lines materialise respectively the spine and the fan separatrices. The right panel shows the co-spatiality between the chromospheric solar ribbons observed at 1600~\AA{} and the location of the separatrix footpoints. The x and y axis units are in arc second defined by the CEA projection.
}
  \label{fig3}
\end{figure*}

In this extrapolated field, using the method of \cite{Demoulin_al94} we find a 3D null-point located around $(x,y) = (632,84)^{\prime \prime}$ and $z=11~\rm{Mm}$. The null point divides the coronal volume in two connectivity domains, the inner and the outer connectivity domain, separated by the fan surface including the null. In each domain a spine is present, which we refer to as the inner (resp. outer) spine for the one confined below (resp. emerging away from) the fan surface. On Figure~\ref{fig3},  we plot separatrix field lines emanating from the null point.  The yellow field lines materialize the dome-like fan separatrix while the dark blue line shows the outer spine that intersects the fan surface at the null point. The part of the spine rooted in the parasitic polarity, the inner spine, is observed on Figure~\ref{fig4}, bottom row. 

The magnetic field extrapolation is performed in a local reference frame. In order to compare the EUV observations with the extrapolated magnetic field, we treat the AIA data in the same way as the magnetic field data used as input of the extrapolation routine, i.e. using Cylindrical Equal-Area projection (CEA). Some of the images are thus presented as if the observer was at the zenith of the center of the selected region.  Because of this CEA projection, 3D structures in the volume such as loops are slightly distorted comparing to the magnetic field lines. However, the morphology of the structures at the low atmosphere level are preserved. Working with a CEA projection and re-projecting the magnetic field model on the plane of sky leads to minimum differences. Since we mainly focus on the comparison between low atmosphere features with the magnetic field,  we obtain a better co-alignement  with the CEA projected data.
Therefore, AIA images are treated similarly to the HMI magnetograms. The small difference of pointing and distortion between the HMI data and the AIA data are compensated by using small variation of the orientation and of the pixel size in order to obtain the best visual fit between iso-contours of the longitudinal magnetic field and the plage observed at 1600 \AA{}. The overall final precision of the co-alignment is of the order of $1-2^{\prime \prime}$. 

By overplotting the separatrix field lines on the chromospheric ribbons (Figure~\ref{fig3}, right panel), we find that the fan footpoints match perfectly (within the co-alignement precision) the location of the quasi-circular ribbon. However, the outer spine footpoint anchors $\simeq 5~\rm{Mm}$ westward of the remote ribbon western extremity. Moreover, the remote ribbon brightens first at the bottom of the inverse-Z shape \citep{Deng_al13}. Considering that the remote ribbons first appears when null-point reconnection starts, the outer spine should initially be anchored at the bottom segment of the remote ribbon and not the upper part as given by the extrapolation. In this case, the extrapolated spine is anchored less than $\simeq 14~\rm{Mm}$ west of the expected spine position. Despite this discrepancy, our NLFFF extrapolation leads to an unequaled overlap between the flare ribbons and the separatrix footpoints compared to previous studies.

In the NLFFF extrapolation by \citet{VemareddyWiegelmann14}, the footpoint of the spine related post-flare loops and the outer flare ribbon are located more than $35~\rm{Mm}$ away from the position of the outer spine footpoint. \citet[][Figure 5]{Yang_al15} showed that the outer spine is about $25^{\prime \prime}$ East of the brightest portion of the outer ribbon. The footpoints of the spine QSL-halo extend strongly westward and the shape present similarities with the elongated outer ribbon. Its position is further south of the actual observed ribbon \citep[see Figure 6 of ][]{Yang_al15}. A fraction of the field lines forming this outer spine QSL-halo are anchored in the outer region between $10^{\prime \prime} - 20^{\prime \prime}$ southward of the outer ribbon. No studies have yet been able to obtain a complete match between all the 3D null-point topological structures with the different flare ribbons. The correspondance of such details of the coronal field models with the flare brightening are revealing of the accuracy in reproducing both the correct current distribution as well as the large scale magnetic field configuration.

The excellent match between the magnetic model and the UV observed features compared to previous studies can be explained by several factors. First, the AR is close to central meridian at the time selected for the NLFFF modelling. Therefore, the vertical component of the field is quite close to the measured line-of-sight one, and limiting the preprocessing to the transverse component, only prevents heavy modifications of the vertical component which is better measured. In addition, the relatively limited preprocessing that we perform, even though it induces some residual forces, avoids large modifications of the distribution of injected currents at the photosphere. Furthermore, the present AR is modelled only a few minutes before a flare event occurs. The real magnetic field within the active region is thus likely not fully force free at the time of our study. Our magneto-frictional approach, while trying to reach a force-free state, does not strictly impose it. The finite force-freeness present in the system may actually enable more realist modelling of the close-to-flaring active region. In combination with the efficiency of the magneto-frictional scheme in injecting properly current, regardlessly of the topological complexity of the field \citep[see][]{Valori_al12}, the careful preprocessing allows a reconstruction of the coronal field that is accurate on different scales simultaneously. In particular, the match between the reconstructed spine position, and associated QSL halo, with the magnetic-model-independant observations of the remote kernel brightening provides a crucial confirmation of the accuracy of our magnetic model. 

By comparison with a potential field extrapolation that we performed (not shown here), the outer spine in the potential field extrapolation connects to the photosphere at a point that is closer to the fan (around $[x,y] = [700,90]^{\prime\prime}$ for the potential model vs. $[x,y] = [720,105]^{\prime\prime}$ for the NLFFF one, as shown left panel of Fig.~\ref{fig3}) and has an associated QSL structure which extends in direction south-east, which is not matching the observation. The injection of currents allowed by the NLFFF model, albeit not introducing currents at the null itself due to its force-free hypothesis, is nevertheless able to change the local and global field to have an improved match of the outer spine  with the remote brightening kernel. In addition, further observational matches discussed below, which are related to the flux rope underneath the fan, are essential nonlinear features which are obviously absent from the potential field model.

 \subsection{Quasi-Separatrix Layers} 
 \label{Q-factor}

Solar ribbons are also observed at QSL footpoints \citep[e.g.][]{Demoulin_al96a,Mandrini_al96,Savcheva_al12}. The regular "cut-and-paste" reconnection regime transfers flux from one connectivity domain to another and leads to ribbon motions perpendicular to the polarity inversion line (PIL) \citep{Bogachev_al05,Aulanier_al12}.  Reconnection across QSLs leads to the flipping/slipping motions of field lines \citep{HessSchindler88,PriestForbes92b}  and causes brightening propagation parallel to the PIL \citep[e.g.][]{Janvier_al13,Janvier_al16}. The critical role of the QSL reconnection in understanding the dynamics of solar flares has been widely demonstrated by observational \citep{Aulanier_al07,Dudik_al14,Masson_al14} and numerical studies \citep{Aulanier_al06,Masson_al09b,Janvier_al13,Savcheva_al16}.
   
\begin{figure*}
 \setlength{\imsize}{0.99\textwidth}
   \includegraphics[width=\imsize,clip=true]{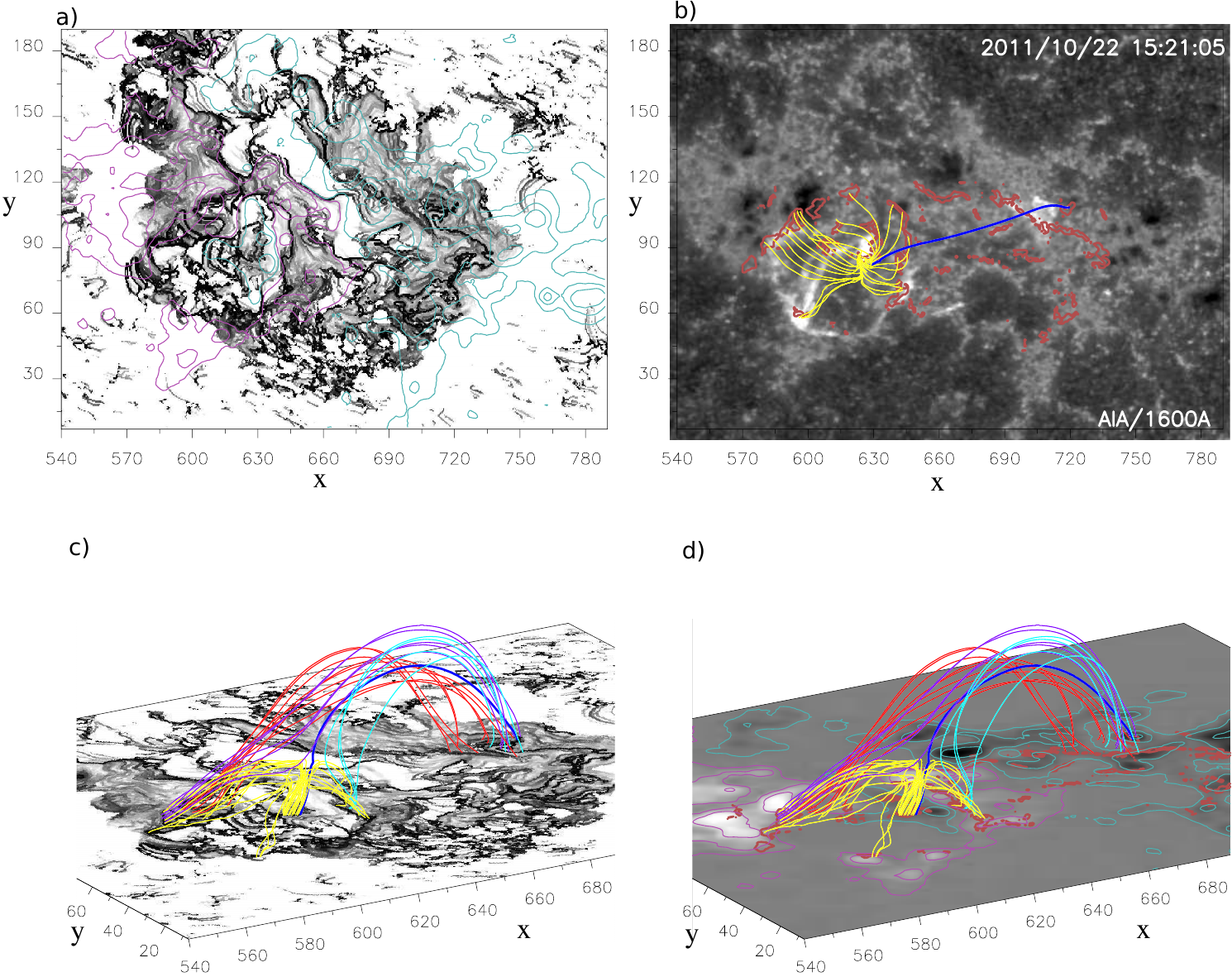}
\caption{QSL's distribution in the domain. Panel a) shows a top view of the Q-factor distribution at $z=0$ between $\rm{Min}(\log Q)= 1$ in white  and $\rm{Max}(\log Q)=  8$ (black) with pink and blue iso-contour of the vertical magnetic field (see Fig.\ref{fig3}). Panel b) shows a top view of the AIA 1600~\AA{} chromospheric ribbons co-aligned with red Q iso-contours such as $\log Q = 6.5$, and the fan and spine separatrix field lines. Panels c) and d) displays the 3D view of the same Q-map grey-scaled and the vertical magnetic field, respectively, with the fan and spine separatrices and the QSL-halo connectivities. The dark blue and yellow field lines shows respectively the spine and fan separatrix field lines,  the red and purple lines belong to the outer-spine QSL halo and connect respectively the north and the east of the fan (see text for details). The x and y axis units are in arc second defined by the CEA projection.}
\label{fig4}
\end{figure*}

The QSL's distribution can be quantified by the so-called squashing factor Q \citep{Titov_al02,Titov07}. It computes the connectivity gradients between one field line and its neighbouring field lines. We use method 3 of  \cite{PariatDemoulin12} to compute the squashing-factor in the full 3D volume. Such method is then applied to the extrapolation described in \S~\ref{NLFFF extrapolation} to compute the squashing factor Q. Figure~\ref{fig4}a shows a top view of the photospheric trace ($z=0$) of the Q distribution in the region. In addition to high-Q region showed by the black lines, we notice a  complex Q distribution. It is a consequence of the non-linear nature of the extrapolation method which, first, maintains enough complexity of the observed photospheric field, secondly, de facto considers a non-force-free equilibrium for the field.

On Figure~\ref{fig4}b we over-plot red iso-contours of high-Q ($\log Q = 6.5$ ) on 1600~\AA{} image. Compared to figure~\ref{fig4}a, the Q values are also  filtered in order to plot only regions where $|B|>50~\rm{G}$ and thus remove the multiple high-Q regions related to weak small-scale or  noise-level magnetic-field measurement. This magnetic field filtering allows us to select the QSLs which are more likely to be associated with the actual flare ribbons. Indeed, the low-lying QSL are unlikely to store large amount of currents and being activated during the flare. Moreover, particle acceleration and ribbon formation are likely to be more important in more intense field regions.  We indeed observe that the plotted high-Q structures are located at the footpoints of the fan surface/circular ribbon (footpoints of the yellow lines), at the inner spine/ribbon (around $x=630^{\prime \prime}$ and $y=90^{\prime \prime}$), as well as at bottom section of the outer spine/ribbon (around $x=705^{\prime \prime}$ and $y=110^{\prime \prime}$). This confirms the results of \citet{Masson_al09b} and \cite{Pontin_al16}, that a QSL-halo surrounds the fan/spine separatrices. Moreover, we notice that the magnetic filtering applied to plot the QSL along the fan shows that those high-Q regions are co-spatial with  the most intense ribbons, R1 ($[x,y]=[585,95]^{\prime \prime}$), R2 ($[x,y]=[595,60]^{\prime \prime}$) \& R3 ($[x,y]=[640,75]^{\prime \prime}$) in Figure~\ref{fig4}b.

Contrary to previous circular flares with elongated and linear inner and remote ribbons, here, they are rather compact. However, the outer spine QSL-halo is actually quite extended, with a length of about $120$ Mm (for $\log Q = 6.5$), with extremities at $[x,y]=[670,120]^{\prime \prime}$ and at $[x,y]=[705,50]^{\prime \prime}$  (see Figure~\ref{fig4}b). The outer ribbon observed at 1600~\AA{} corresponds to a small fraction of that extended outer spine QSL-halo. In the image at 131~\AA{}, this very extended QSL corresponds to brightenings observed prior to the studied flare (not shown here).  The 1600~\AA{} outer ribbon is however located close to the regions having the highest Q value of that extended outer-spine QSL-halo (Figure~\ref{fig4}b). Furthermore the remote ribbon has an inverse-Z shape \citep[cf.][which extensively discuss the shape and dynamics of the outer ribbon]{Deng_al13}. This inverse-Z shape is also observed in the QSL distribution around $[x,y]=[705,105]^{\prime \prime}$ (Figure~\ref{fig4}a), precisely at the location of the observed outer ribbons. This hints that the shape of the flare ribbons are morphologically controlled by the QSLs as discussed in \citet{Masson_al09b}. Conversely, this is an additional confirmation of the quality of the extrapolation since not only the location of the outer ribbon is reproduced  but the entire QSL halo surrounding it is correctly represented.

In order to determine the QSL's connectivity in the flare domain, we plot field lines from footpoints anchored in region of high Q. On figures~\ref{fig4}c \&~\ref{fig4}d we show 3 groups of field lines that connect the outer ribbon and the quasi-circular ribbon. Those group of field lines materialize the QSL-halo surrounding the fan and spine separatrices. The North-East part of the circular ribbon (R1, Figure~\ref{fig1}) is connected to the upper part of the inverse-Z ribbon by the purple field lines and to the bottom part of the remote ribbon by the red field lines. The outer spine is also connected to the South West part  of the fan (R3) by the light blue field lines. Studying the QSLs connectivity allows us to identify in details the section of the QSLs associated with the UV ribbons observed at 1600~\AA{}.

\subsection{Flux rope under the fan}
\label{Flux rope}
 
\begin{figure*}
 \setlength{\imsize}{0.99\textwidth}
   \includegraphics[width=\imsize,clip=true]{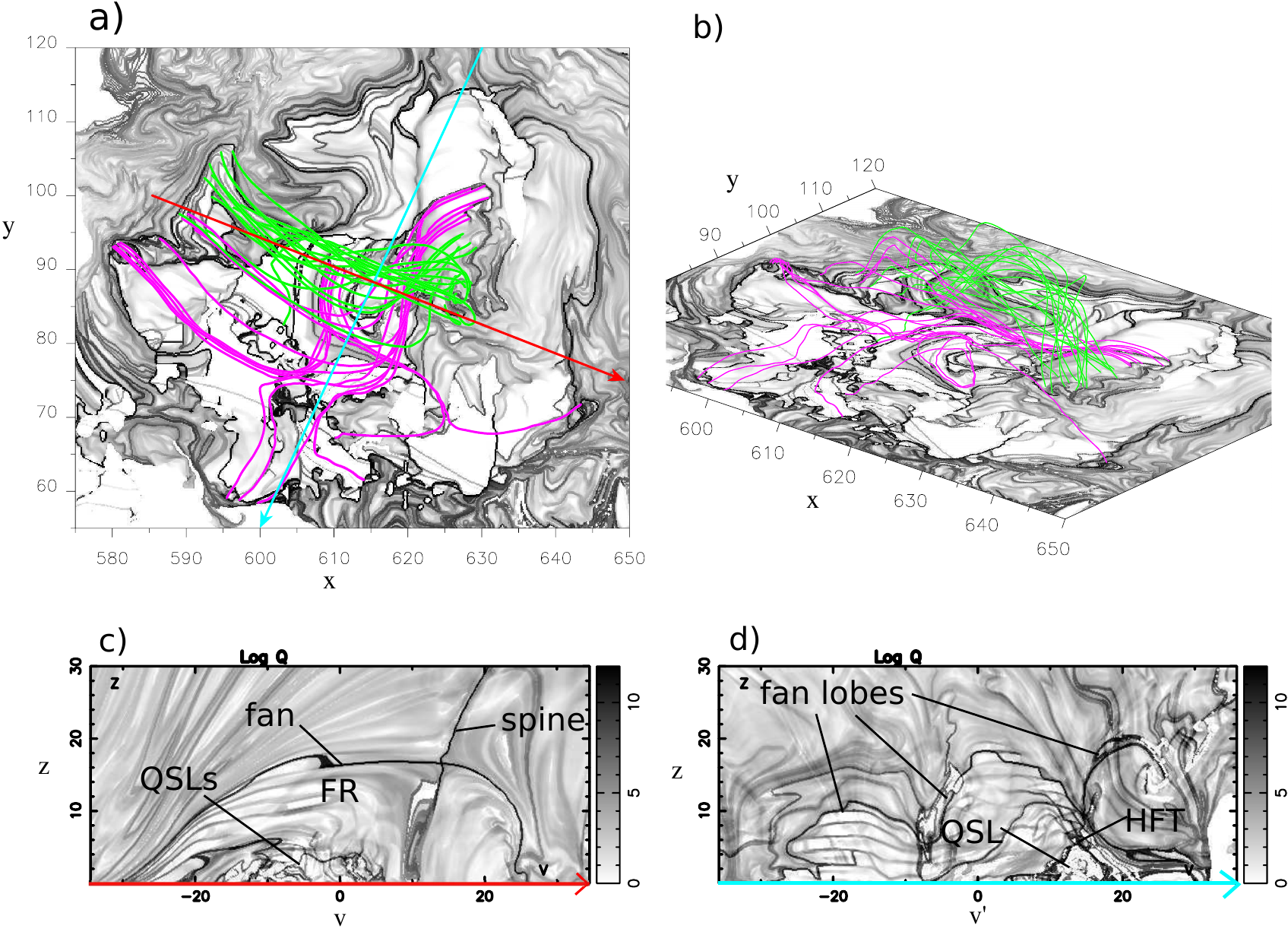}
\caption{Q-factor distribution in the inner connectivity domain. Panel a) and b) shows a top and 3D view of a zoom of the distribution of the Q-factor at $z=0$  in the parasitic polarity region enclosed below the fan area, with $\log Q= 1$ in white and $\log Q= 12$ in black. The pink lines show low-lying internal QSLs confined below a green flux rope.  Panels c) and d) display vertical 2D cuts, along the red and blue arrows plotted on panel a) respectively, of the Q-factor computed from $z= 0$ to $z=30~\rm{Mm}$. The Q-factor is grey-shaded with white and black respectively showing $\log Q= 0$ and $\log Q= 12$. The x and y axis units are in arc second defined by the CEA projection.}
\label{fig5}
\end{figure*}

Enclosed in the fan surface, we identify small scale structures of high-Q (Figures~\ref{fig4}a and \ref{fig4}c). Figure~\ref{fig5}a and ~\ref{fig5}b, present a zoom of the fan region respectively with a top and a 3D view. By plotting field lines from those high-Q regions we found  a twisted flux rope (green field lines). This structure is clearly helical and the number of turns is of the order of the unity. This flux rope connects the north part of the circular ribbon R1 ($[x,y]\sim[595,105]^{\prime \prime}$) and  QSL-halo of the inner spine ($[x,y]\sim[630, 90]^{\prime \prime}$), i.e., the inner ribbon. The footpoints of the flux rope correspond to two regions of high EUV emission which indicates a preeminent role in the dynamics of the flare. While flux ropes have commonly been observed confined below null-point dome in extrapolation of major eruptive flares (see~\S~\ref{Introduction}), \cite{Wang_al15} and \cite{Liu_al16} found a flux rope for a confined flare in a bipolar active region. In our event we also found a flux rope, but this time it is confined below a null-point (tripolar active region).  This  opens a challenging question on the universality of flux ropes for impulsive flares, whether they are eruptives or not, independently of the topology.

Panels c and d of Figure~\ref{fig5} present two vertical cuts of the Q distribution, in the direction along the flux rope (Figure~\ref{fig5}c) and perpendicular (Figure~\ref{fig5}d) to it, respectively indicated by a red and a cyan arrow in Figure~\ref{fig5}a. Similarly to \citet{Sun_al13,Yang_al15,VemareddyWiegelmann14}, these cuts allow us to study the null point structure and the connectivity in the fan dome. In Figure~\ref{fig5}c the section passes close to the null point, and we clearly identified the null and its separatrices (see labels on Fig~\ref{fig5}). In the local framework [v,z] of Figure~\ref{fig5}c, the null is located at the intersection of two very high Q lines (in black) at $[v,z]=[13,16]$. This 2D cut, shows the 2D version of a null-point topology. The fan separatrix corresponds to the arc-shape black line, and the outer, respectively the inner, spine corresponds to the black line of high Q almost vertical and extending upward, respectively downward, from the null. The 2D  perpendicular cut (Figure~\ref{fig5}d) is further away from the null and the fan present a multi-lobe morphology instead of a smooth arc-shape structure. While a smooth dome shape of the fan is standard for idealized null points \citep{Pariat_al09,Masson_al12a,Sun_al13,Pontin_al16} the fan surface in real solar events can present a complex 3D shape. 

The quasi-connectivity domain containing the flux rope is bounded at the top by the fan surface and at the bottom by a network of low-lying QSLs (see labels on Figure~\ref{fig5}c \& d). The flux rope indeed corresponds to an independent quasi-connectivity domains bounded by QSLs. Part of the low-lying QSLs forms an hyperbolic flux tube \citep[HFT][]{Titov_al03}, e.g. located at $[v',z]=[14,6]$ in Figure~\ref{fig5}d. HFTs are characteristic features associated with flux ropes which have been predicted theoretically and clearly identified in numerical models and observations \citep{Aulanier_al10,Savcheva_al12a,Savcheva_al12b,Savcheva_al15,Janvier_al13,Zhao_al14}
. The field lines belonging to  this HFT are plotted in pink on Figure~\ref{fig5}a and~\ref{fig5} b. These field lines  bound the flux rope from below and connect some spots of the inner-spine QSL-halo with several distinct point of the inner fan. As one can see on the 2D cuts, the observed HFT is far more complex than in toy models \citep[e.g. ][their figures 4 and 7]{Aulanier_al06}.


\section{Impulsive phase}
\label{detailed flare dynamics}
 
In order to investigate the dynamics of the flare, we co-align the AIA multi-wavelength images and the magnetic field data derived from the extrapolation. This allows us to identify the topological elements (\S~\ref{Q-factor}) associated with the plasma emission along the observed  EUV flare loops. Field lines present in the extrapolation of the 15:00UT magnetogram are expected to be present in a very similar state until major changes of the photospheric magnetic field occur. For this reason, even though there is $10$ to $20~\rm{min}$ time difference between the magnetogram and the EUV signatures, it is possible to use the NLFFF extrapolation field lines to model the pre-flare configuration that leads to the observed post-flare loops. To do so we overplot field lines of the topological elements identified in \S~\ref{magnetic topology} on AIA images. To minimize the ambiguity on the thermal properties, we select the AIA channels with well defined temperature response function to diagnose the hot, warm and cold plasma emission \citep{Reale10,Guennou_al12a}. Therefore, we ignore the 335\AA{} line. However, we use the 131\AA{} line to track the hot plasma emission. Even though this line has 2  contributions to the line width they are well separated and are sensitive to a cold ($T\simeq 10^{5.6}~\rm{K}$) and a hot ($T\simeq10^{7}~\rm{K}$) component. According to the AIA observations that are described below, the hot component of the line is likely to be the majority emission at 131\AA{}. We also prefer to use the 171\AA{} line ($T=10^{5.8}~\rm{K}$) to diagnose the cool emission instead of the 193~\AA{}  that has a warm ($T\simeq10^{6.1}~\rm{K}$) and a hot peak ($T\simeq 10^{7.3}~\rm{K}$). Finally, we use 211\AA{} line peaking at a warm temperature ($T\simeq 10^{6.3}~\rm{K}$).

\subsection{Reconnection along the Hyperbolic Flux Tube}
 \label{internal QSL}
 
\begin{figure}
 \setlength{\imsize}{0.49\textwidth}
   \includegraphics[width=\imsize,clip=true]{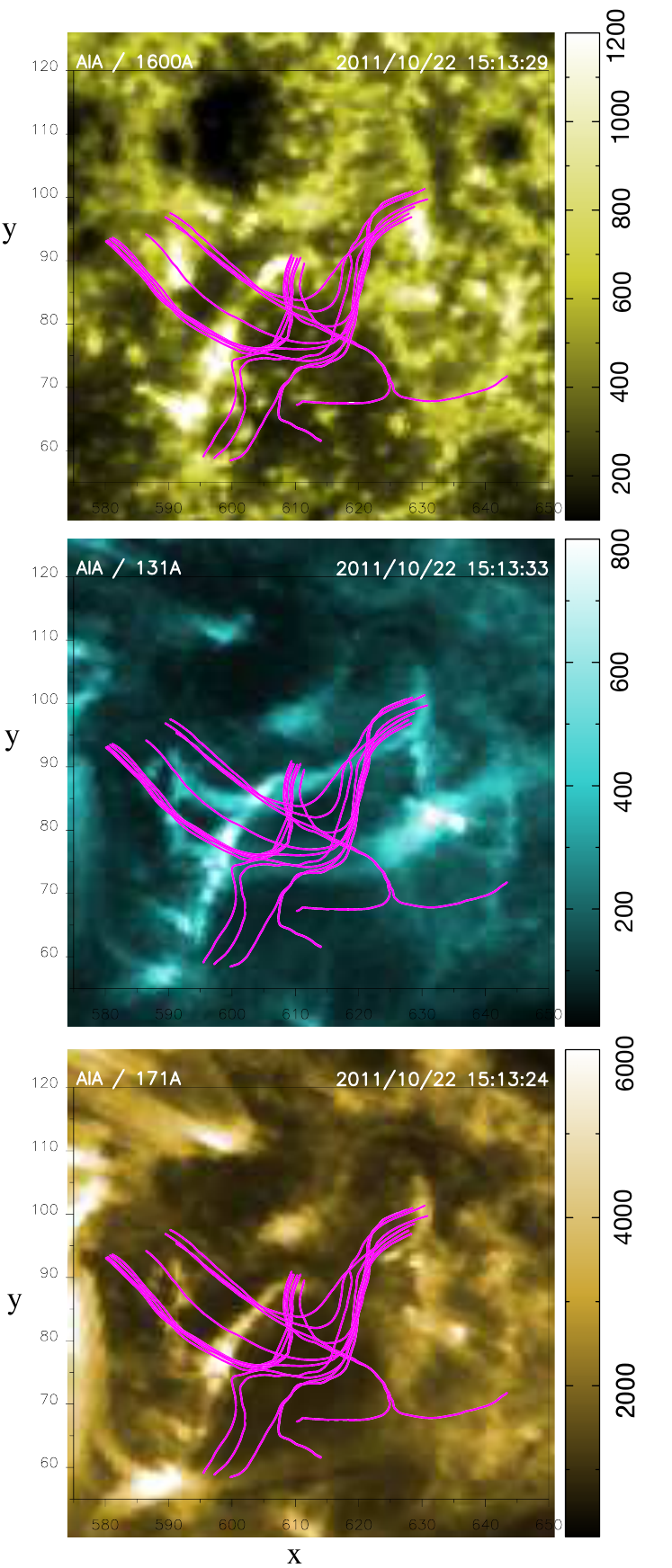}
\caption{AIA images co-aligned with the magnetic field extrapolation zoomed in the region where the pink HFT related QSL field lines are observed. The CEA projection has been used for the co-alignement. From top to bottom, the EUV images are at $1600$~\AA{}, $131$~\AA{} and  $171$~\AA{} and  showing multi bright strands and kernels. The field of view for the magnetic field data is $x=[580,650]^{\prime \prime}$, and $y = [55,120]^{\prime \prime}$ in the CEA projection frame. The field of view of the AIA images is slightly larger.}
\label{fig6}
\end{figure}

 At the beginning of the flare, while the null-point reconnection has not started yet (no outer spine brightening are observed), some brightenings are observed inside the quasi-circular ribbon (see \S~\ref{UV ribbons}). Figure~\ref{fig6} shows those brightenings detected at 1600~\AA{} , 131~\AA{} and 171~\AA{} at 15:13~UT. On each AIA image, we overplot the HFT pink field lines. First, their footpoints are perfectly co-spatial with the bright kernels observed at 1600~\AA{}. Second, we identify two bright loops (or strands) in the hot (131~\AA{}) and in the warm (171~\AA{}) lines (see online animation : movie 2 and 3 and Figure~\ref{fig6}). Those strands connect the 1600~\AA{} bright kernels located at the footpoints of the HFT pink field lines and have an overall shape very similar to the HFT pink field lines. Contrary to the bright kernels at 1600~\AA{}, those bright loops are not completely co-spatial with the HFT pink field lines. The observed shift on Figure~\ref{fig6} between the loops and the magnetic field lines is a direct consequence of the applied CEA projection (\S~\ref{Null point topology}). The spatial correspondence between the kernel and the HFT's footpoints suggests that reconnection across those QSL is most certainly responsible for the EUV emission observed with AIA very early during the flare, prior to the onset of reconnection at the null-point (see \S~\ref{UV ribbons}). The presence of bright loops connecting those bright kernels and displaying a shape very similar to the HFT pink field lines argues strongly in favour of reconnection inside the HFT and its associated QSL, leading to the EUV emission along the loops and at the loop footpoints, confirming the results of \cite{Reid_al12}. Furthermore, those brightenings last over the entire flare indicating that reconnection is still occurring at the HFT.

\begin{figure}
 \setlength{\imsize}{0.49\textwidth}
   \includegraphics[width=\imsize,clip=true]{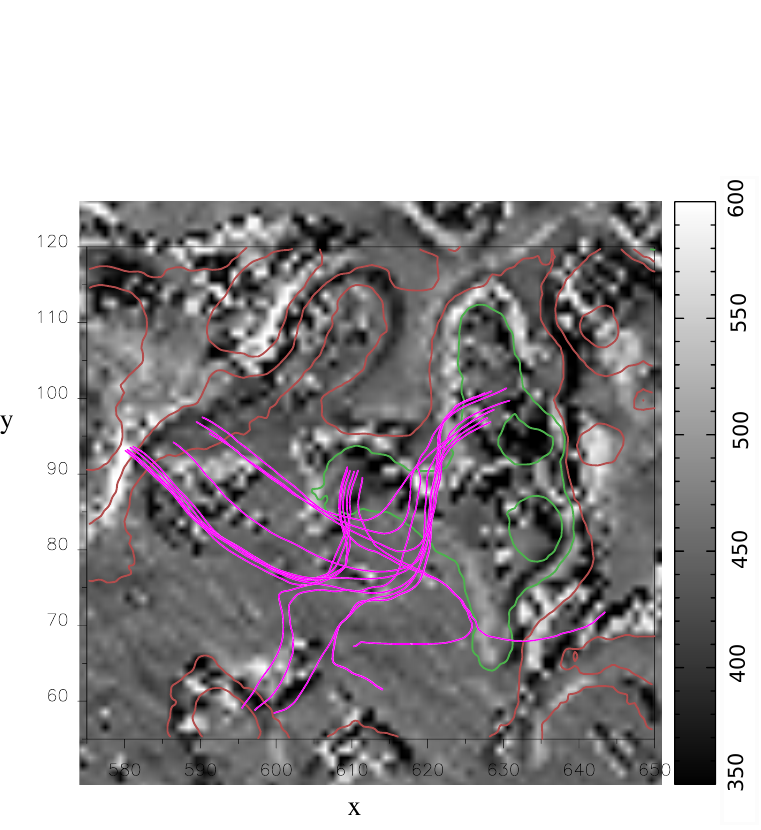}
\caption{ Top view of the photospheric map of the vertical current resulting from the NLFFF extrapolation from the HMI vector magnetogram (colored in greyscale). The color-scale ranges from $350\times1.6\times10^{-6}~A.m^{-2}$ to $600\times1.6\times10^{-6}~A.m^{-2}$. Red and green iso-contours correspond to positive and negative $B_{los} = \pm 40, 250, 750~\rm{G}$, respectively. The pink field lines correspond to the HFT related QSL field lines. The field of view for the magnetic field data is $x=[580,650]^{\prime \prime}$, and $y = [55,120]^{\prime \prime}$ in the CEA projection frame. The field of view of the vertical current image  is slightly larger.
} 
\label{fig7}
\end{figure}

 While the photospheric currents are usually cleaned by the pre-processing of the magnetogram for numerical stability reason, we were careful to keep the most significant part of  the photospheric electric currents in our pre-processing and extrapolation method (\S~\ref{NLFFF extrapolation}). On Figure~\ref{fig7} we focus on the fan region and we show the vertical current density at the photosphere. We overplot the HFT's pink field lines in order to address the origin of the bright kernels and loops observed inside the fan. The footpoints of the HFT pink field lines, which is a coronal structure, are located in region of high photospheric vertical current density. Vertical current density observed at the solar surface are the photospheric trace of the currents flowing through magnetic structures. The presence of strong currents located at the HFT footpoints is a strong evidence that the HFT magnetic field carries electric currents. By definition, the HFT is a region with a high Q values, i.e. a preferential site for magnetic reconnection \citep{Aulanier_al05}. The presence of currents in the HFT is  undeniably the best observational evidence that magnetic reconnection occurs in the HFT and the observed bright loops and kernels inside the fan result from the HFT magnetic reconnection.  While  HFTs are often found in observations \citep[e.g.][]{Savcheva_al15}, to our knowledge, this is the first time that magnetic reconnection along an HFT has been clearly identified in an observation of a non-eruptive compact flare.

 \subsection{Null-point reconnection}
 \label{null point}

\begin{figure}
 \setlength{\imsize}{0.49\textwidth}
   \includegraphics[width=\imsize,clip=true]{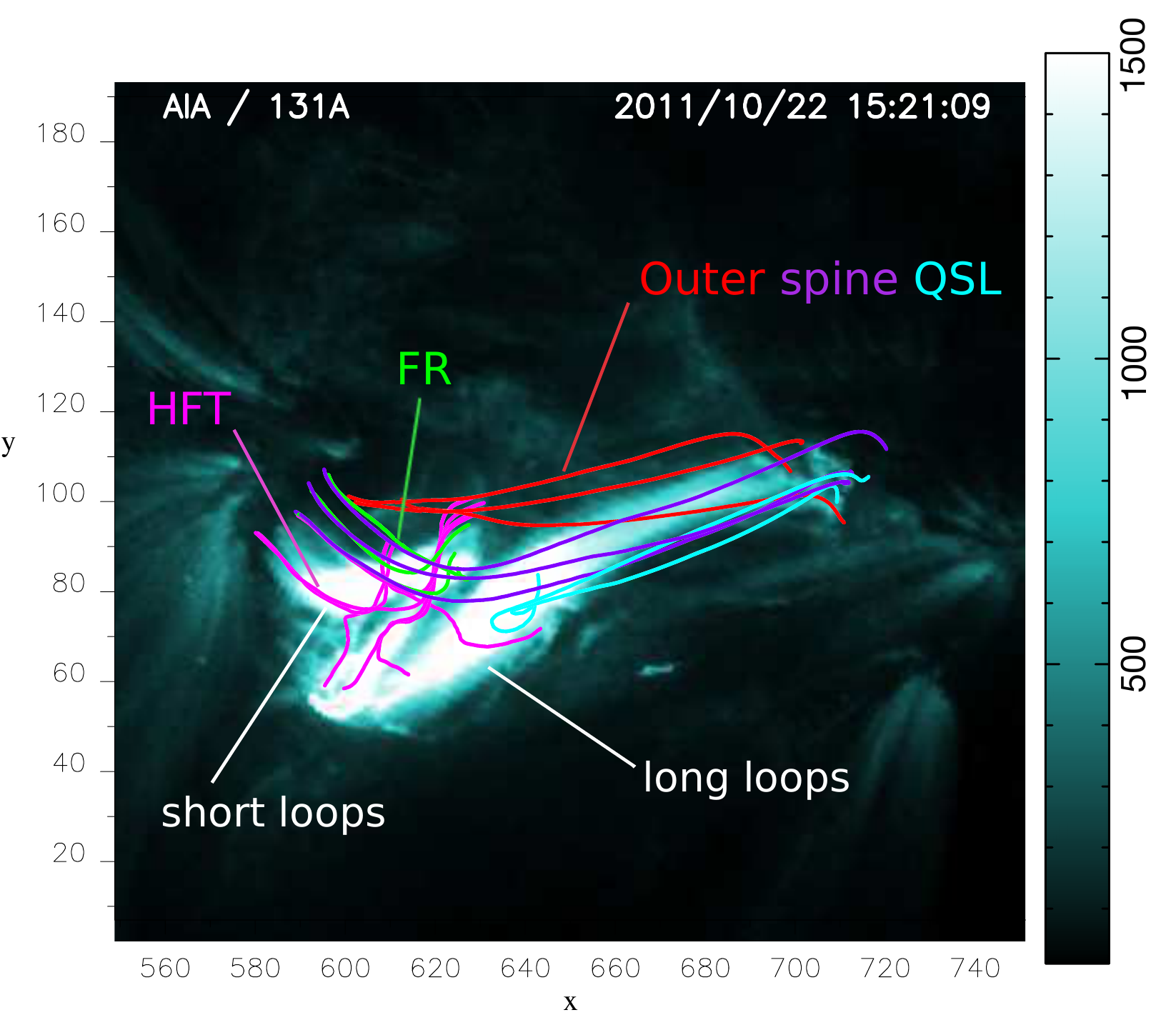}
\caption{The association between the EUV post flare loops and the topological elements : the 131~\AA{} image at the time of the main phase of the flare (15:21~UT) on which is overplayed the green field lines corresponding to the flux rope structure, the pink lines showing the HFT related QSL field lines, the red and purple lines represent the outer-spine QSL-halo, the light blue field lines show the field connecting the outer spine and R3. The CEA projection has been used for the co-alignement and the x and y axis units are in arc second defined by the CEA projection.}
\label{fig8}
\end{figure}

\begin{figure*}
 \setlength{\imsize}{0.99\textwidth}
   \includegraphics[width=\imsize,clip=true]{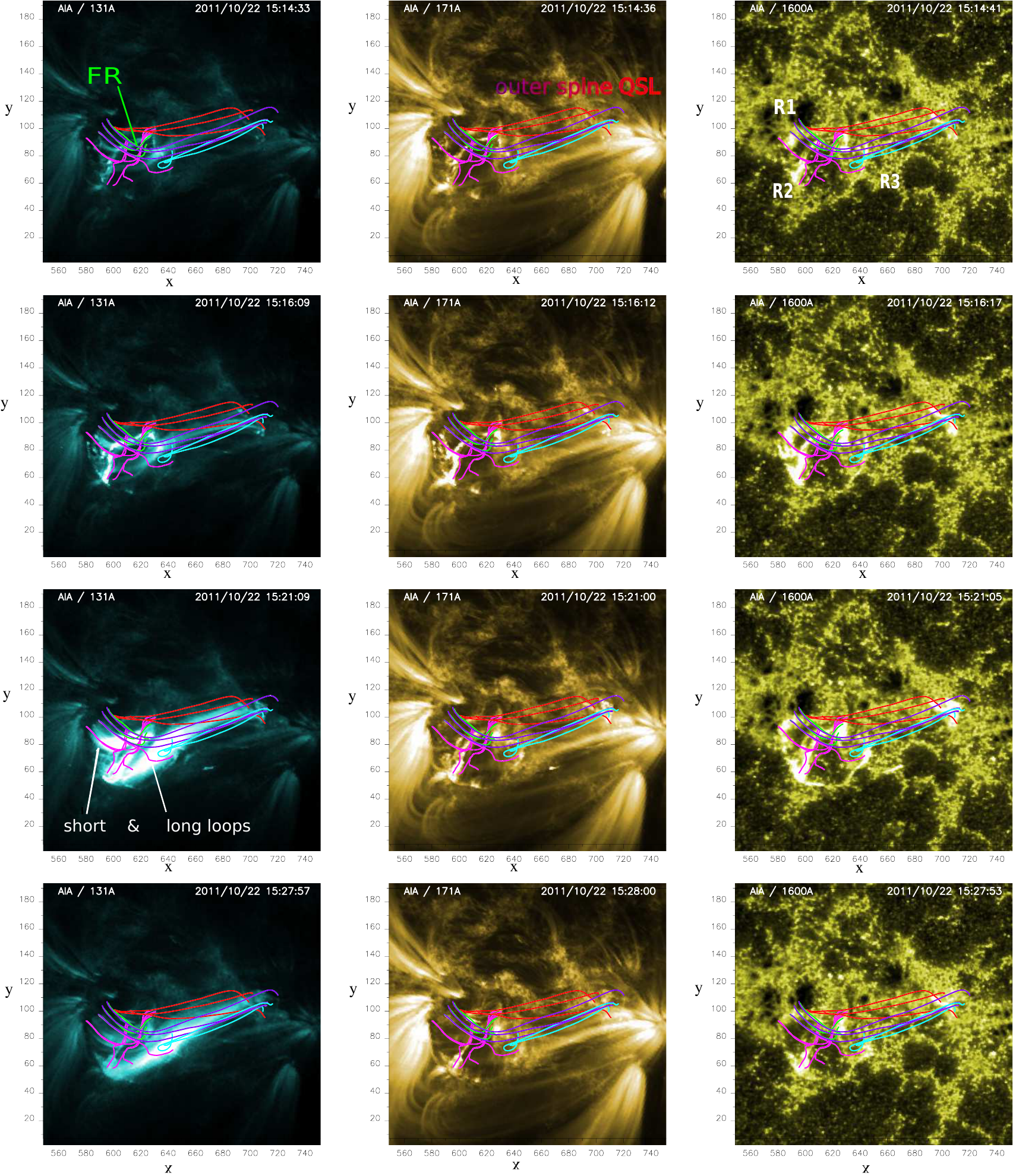}
\caption{Time sequence of the AIA images during the main reconnection episode of the flare. Each column corresponds to an AIA channel observing the flare, from left to rigth: $131$, $171$ and $1600$~$\AA{}$ and each row displays a time during the main null-point reconnection episode, from top to bottom: 15:14:21~\rm{UT}, 15:16:09~\rm{UT}, 15:21:00~\rm{UT} and 15:27:57~\rm{UT}. The field lines over plotted are coloured as described in the main text and in Figure~\ref{fig8} caption. The CEA projection has been used for the co-alignement. The field of view for the magnetic field data is $x=[550,750]^{\prime \prime}$, and $y = [10,190]^{\prime \prime}$.
}
\label{fig9}
\end{figure*}

 Figure~\ref{fig8} shows the 131~\AA{} AIA image at the time of the flare (15:21~UT) on which we overplot only few field lines of each topological elements: the HFT (pink), the flux rope (green), the upper (purple) and bottom (red) outer spine QSL field lines. For clarity, this figure should be used as a reference in this section. After the HFT reconnection phase, the system enters its main flare phase. The time sequence is shown on Figure~\ref{fig9}. It includes the main flare episode during which solar ribbons form, extend  and intensify, and the beginning of decaying episode when the brightenings start to faint (15:24~UT). On each panel we plot the same field lines as in figures~\ref{fig8} characterising each identified topological elements (see labels on figure~\ref{fig8}).
On the right column of Figure~\ref{fig9}, we associate the field lines with the UV ribbons described in \S~\ref{UV ribbons} (and indicated on Figure~\ref{fig1}):

\begin{itemize}
\item the footpoints of the green flux rope connect  R1 and the inner spine ribbon (see \S~\ref{Flux rope}); 
\item  the red outer-spine QSL field lines  connect the north part of the circular ribbon (R1) and the south east part of the inverse-Z outer spine ribbon (see \S~\ref{Q-factor}); 
\item the purple outer-spine QSL field lines connect R2 and the north west part of the inverse-Z outer spine ribbon (see \S~\ref{Q-factor}) 
\item R3 is located at the footpoints of the yellow fan field lines and is connected to the outer spine ribbon by the light blue field lines. 
\item As discussed in section~\ref{internal QSL}, the QSLs of the HFT (pink field lines) connect the inner spine ribbon mainly with R1 and R2 (Figure~\ref{fig6}). 
\end{itemize}

In addition to the association of the field line footpoints and the photospheric brightenings, we use the other AIA channels to associate the field with post flare-loops. Short and long bright loops (see labels on figure~\ref{fig8}) first appear at 15:14~UT at 131~\AA{} (first column on Figure~\ref{fig9}) but not in the warm (171~\AA{} and 211~\AA{})  line until the late EUV phase (see \S~\ref{EUV late phase}). The short loops are located beneath the fan surface and connect $[x,y]\simeq[580, 90]~\rm{^{\prime\prime}}$ and $[x,y]\simeq[620, 90]~\rm{^{\prime\prime}}$, while the longer loops connect the eastern part of the circular ribbon  ($[x,y]\simeq[580, 50]~\rm{^{\prime\prime}}$) and the remote ribbon ($[x,y]\simeq[700, 100]~\rm{^{\prime\prime}}$). Based on the magnetic field distribution, the short loops  match the green flux rope and originate from magnetic reconnection of the overlying field across the null-point and the associated QSL-halo. The long loops are co-spatial with the purple outer spine QSL-halo and correspond to field lines originally beneath the fan that have reconnected across the null point and the associated QSL-halo.
Since the post-flare loops are observed after reconnection occurred,  the bright loops and the extrapolated pre-flare field lines do not overlap perfectly. Nonetheless, the good agreement between the 131~\AA{} loop footpoints, the chromospheric ribbons and the reconstructed magnetic field lines (see Figure~\ref{fig8} \& ~\ref{fig9}) strongly indicates that the flux rope and the outer QSL are the principal magnetic fluxes involved during the flare.

Since solar ribbons trace the footpoints of the reconnected flux, their spatial evolution provides informations about the dynamics of the flux transfer during the flare. Indeed, a displacement of the solar ribbons perpendicular to the polarity inversion line is very likely to indicate a flux transfer between two distinct connectivity domains, while a  motion parallel to the PIL suggests a continuous change of connectivity in a single connectivity domain.
As described in \S~\ref{UV ribbons} and seen on the 1600~\AA{}  high cadence movie 1 (see online material), the global spatial evolution of the circular ribbons displays a counterclockwise propagation. In addition, R1 moves northward perpendicularly away from the PIL,
%
and indicates a flux transfer from the inner to the outer connectivity domain in that portion of the fan.

 The displacement of R2 and R3 parallel to the circular PIL suggests a slipping reconnection process. Slipping reconnection that leads to brightening propagation along the PIL is an intrinsic property of magnetic reconnection episode in a null-point topology. It has indeed been widely showed observationally \citep{Masson_al09b,Wang_al12,Sun_al13,Yang_al15} and numerically \citep{Masson_al09b,Masson_al12a,Pontin_al11,Pontin_al13}. These motions occur simultaneously with the extension along the inverse-Z shape of the outer spine ribbon studied in detail by \cite{Deng_al13} which are most likely caused by a slipping reconnection regime across the outer spine QSL-halo in the null vicinity. 
In the meantime, the two 131~\AA{} hot loops exhibit an apparent slipping motion along the quasi-circular ribbon in the counterclockwise direction (see movie 2). This motion is synchronised with the counterclockwise propagation of the chromospheric brightening along the circular ribbon at 1600~\AA{}. The spatial association of those "slipping post-flare loops" with the green flux rope and the outer-spine QSL-halo suggests that the slipping motion of the post-flare loops is the observational consequence of magnetic reconnection across the QSLs (green flux rope and the outer-spine). Both flare loops exhibit a southward motion which is in agreement with the displacement of R1 and R2 along the quasi-circular ribbon.


The synchronous evolution of the short and the long hot loops indicates a common origin. For a 3D null-point there is only 2 distinct connectivity domains and not 4 as in the `2D null-point' picture. Therefore, we do not expect to exchange only flux between two `opposite' domains as pictured in 2D, but between the connectivity domains enclosed below the fan surface and outside of it \citep[e.g.][]{WyperDeVore16}.
 A plausible scenario that could explain the appearance around 15:14~UT of the two flare loops as well as the displacement of  R1 and R2 along the quasi-circular ribbon consists of the exchange of connectivity between the green flux rope and the red and purple outer QSL field lines at the null-point. One footpoint of the flux rope anchored at the fan becomes connected to the outer spine ribbon, forming the long flare loops, and the footpoints of the outer QSL field lines initially connecting  R1 and R2 to the outer ribbon, now connects R1 and R2 to the inner spine ribbon. The new magnetic loops belong to the inner domain forming the short post-flare loops. Initially (at 15:14~UT) the footpoints of the long loops are anchored in the bottom part of the outer spine ribbon (first column on Figure~\ref{fig9}) suggesting that the flux reconnecting in the first place with the flux rope corresponds to the red field line.  After being newly formed by null-point reconnection between the flux rope and the overlying field, the long flare loops are slip-reconnecting across the outer-spine QSL-halo and the short flare loops across the QSL structures surrounding the inner spine as well as the flux rope. In the meantime, the conjugate footpoints of the long loops, anchored at the outer spine ribbon also go under slipping reconnection. This leads first to the extension of the bottom of the outer-spine ribbon and then to the formation of the upper part of the inverse-Z shape \citep{Deng_al13}.

During the flare, R3 does not drift as much as R1 (see movie 1). According to the magnetic field configuration, R3 is located at the footpoints of  yellow field lines enclosed below the fan and at the light blue field line footpoints. The absence of observed post-flare loops connected to R3 prevents us to identify which magnetic fluxes reconnect and lead to R3 formation.  However, it is reasonable to argue that the yellow field lines, initially connecting the inner spine and R3, reconnect with the overlying field, connecting R2 and the remote ribbon. The two resulting fluxes connect R2 to the inner ribbon and the remote ribbon to R3, similarly to the light blue field line. In this scenario, the new flare loops would be hidden by the short and long loops at 131~\AA{}, but would appear later in other channels during cooling episodes, as we do actually observe (see~\S~\ref{EUV late phase}).

Compared to previous null-point eruptive flare studies \citep{Sun_al13,Liu_al13}, our event presents a major difference in the post-flare loops location due to its non-eruptive nature. For an eruptive flare, the reconnection at the null-point aims to remove the confining flux from above the flux rope, creating post-flare loops closing down in the side lobes \citep[][]{Antiochos_al99,Sun_al13}, i.e. below the fan and below the outer spine. During our non-eruptive event, post-flare loops are not observed below the fan, but above the outer spine, i.e., magnetic flux overlying the null-point is increasing, instead of decreasing as expected for a breakout-type eruption. 

The above interpretation of the flare dynamics is able to relate the different elements of the magnetic field topology, as reconstructed via NLFFF modelling from photospheric observations, with the observed time sequence of brightening in practically all AIA channels. We believe that we have identified the main players of the process (the HFT and flux rope system within the large scale null point structure) and interpreted a possible generation sequence of the different observed flare phases.  However, this interpretation remains hypothetical, mainly because observations do not help constraining the flux rope dynamics enough. A proper validation of the above explanation would require a fully fledged MHD simulation that has our NLFFF extrapolation as an initial condition, and evolved under data inspired, or constrained, photospheric flows.
 
 
 \section{EUV late phase}
 \label{EUV late phase}
 
After the main impulsive phase of the flare we identified 3 other peaks in the EUV emission (see \S~\ref{thermal evolution} and Figure~\ref{fig2}) which we consider as late EUV phases.  
 Each of the three local maxima at 335~\AA{} is followed by a peak in colder channels at 171~\AA{} and 211~\AA{}. Each of those three episodes are sequentially organised such as the EUV light curves  peak successively in 211~\AA{} (purple curve on figure~\ref{fig2}), 193~\AA{} (not shown here) and 171~\AA{} (yellow curve), i.e. lines ranging from warm to cold temperatures. This sequential evolution has already been noticed for eruptive X-class circular ribbon flares \citep{Dai_al13,Liu_al13,Sun_al13}. If those EUV late phases are caused by a cooling process of post-flare loops, we expect that the emitting loops associated with the first EUV late phase are shorter than the loops associated with the last episode \citep{Reale10}. 
 
\begin{figure*}
 \setlength{\imsize}{0.79\textwidth}
   \includegraphics[width=\imsize,clip=true]{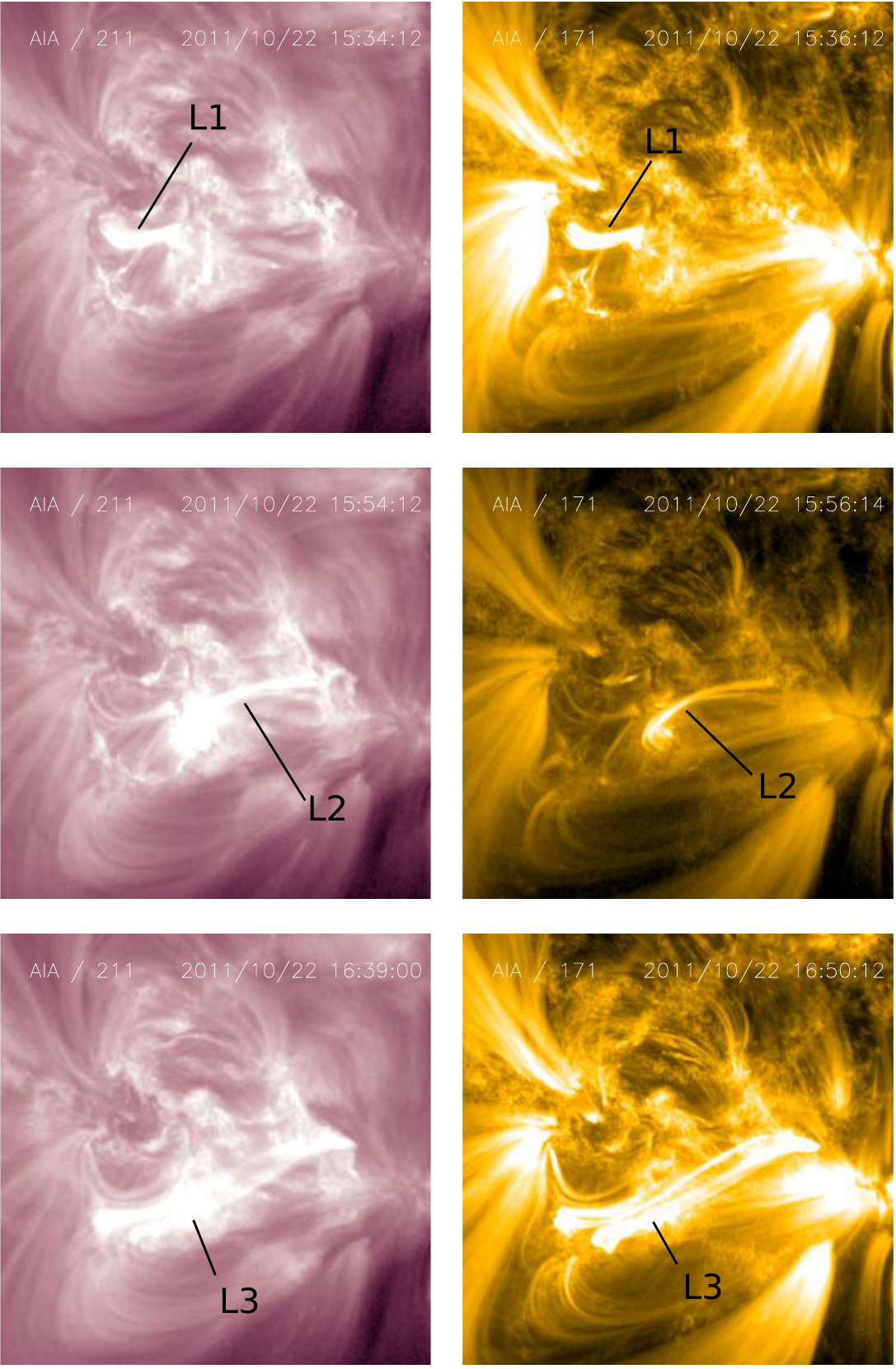}
\caption{Time sequence of AIA images showing one time for each EUV late episode. From top to bottom, the rows show AIA images at $15:34~\rm{UT}$, $15:54~\rm{UT}$ and $16:30:00~\rm{UT}$ . Each column corresponds to an AIA channel observing the flare, from top to bottom: $131$, $211$ and $171$~$\AA{}$. The images are in the observer view and the field of view is the same than figure~\ref{fig1}}.
\label{fig10}
\end{figure*}

As in \S \ref{null point}, we associate the flare loops with the EUV late phase post-flare loops at 15:34~UT, 15:54UT and 16:39 - 16:50~UT (Figure~\ref{fig10}). Contrary to \cite{Woods_al11}, we find that the two hot loops emitting at 131~\AA{} (figure~\ref{fig8} and the first column, third panel on figure~\ref{fig9}) are co-spatial with the loops observed at 171 and 211~\AA{} (Figure~\ref{fig10}) and must therefore have the same origin \citep{Liu_al13}. The first EUV phase correspond to the L1 post-flare loops associated with the green flux rope (first row, figure~\ref{fig10}). The second peak originates from the flare loops, L2, connecting the remote outer ribbon and R3 (middle row, figure~\ref{fig10}), as the light blue field lines do (cf Fig.~\ref{fig8} and ~\ref{fig9}).  Those post-flare loops could coincide with the connectivity exchange between southern fan flux with the outer magnetic flux as described in \S~\ref{null point} to explain R3 formation. Finally, the last peak between 16:39 and 16:50~UT~UT corresponds to the longest loops (L3) which connect the outer spine ribbon to R2 and R3, i.e. the upper outer QSLs (bottom row, figure~\ref{fig10}) .

 The association of the 3 EUV late phases with different sets of flare loops suggests a correlation between the time delay of the EUV late phase and the length of the emitting loops. In our event, the flux rope related flare loops are the shortest (length estimation $L_{L1} \simeq ~35~\rm{Mm}$) and brighten first, $\simeq 16~\rm{min}$ after the maximum of the main episode. Then, the flare loops associated to the second peak are longer ($L_{L2}\simeq 60~\rm{Mm}$) and start to emit $\simeq 36~\rm{min}$ after the main episode peak. Finally, $\simeq 81~\rm{min}$ after the main episode, the last peak indicates the cooling of even longer loops ($L_{L3}\simeq 100~\rm{Mm}$) that connect the outer spine ribbons and R2 \& R3. For each peak, i.e., each post-flare loops, a time delay is observed between the local maxima of the light curves : The cooler line (171) peaks after the warmer one (193), which peaks after the hottest one (211) suggesting a cooling process of plasma-filled loops during those 3 episodes. Those time delay are observed on Figure~\ref{fig2}. Figure~\ref{fig10} displays the AIA images at the different time of the EUV late phase in order to highlight the association of the magnetic field with the loops at their maximum of emission. The spatial location of the emitting loops associated with temporal evolution of the EUV channels strongly indicate that the 3 EUV late phases observed during this event originate from the cooling of post-flare loops by radiative loss process.

 According to the dynamics of the flare inferred from the AIA images combined with  an highly detailed topological analysis (\S~\ref{null point}),  those post-flare loops, i.e. the EUV late phase, result from a single reconnection episode occurring at the null-point. \cite{Sun_al13} have a similar conclusion, but instead of 3 set of post-flare loops they only have 2 associated with the EUV late phases. This difference may be explained by the nature of the flare. Contrary to our confined event, \cite{Sun_al13} studied an eruptive one. In addition to the null-point reconnection there is the 'flare reconnection' below the erupting flux rope which occurs during the main phase of the flare. The resulting post-flare loops (A1-loop in their Figure 1) are very short and their emission contribute to the impulsive phase of the flare, but not to the late EUV phase.  Even though our result indicate a cooling process to explain the EUV late phase, we can not rule out that additional reconnection episodes may occur later in the flare but are hidden by the post flare loops emission. Our results also differ from previous studies concluding that the EUV late phase were caused by several reconnection episodes \citep{Dai_al13, Liu_al13}. However, \cite{Li_al14} revisited 4 events with EUV late phase by analysing their magnetic topology.  They found systematically a null point topology, i.e., magnetic configuration involving loops of very different length.

 \section{Discussion and Conclusion}
 \label{conclusion}
 
We present a comprehensive study of a confined circular flare observed by SDO on October 22, 2011 which extends the study from \cite{Deng_al13} focusing on the spine dynamics. By combining a detailed topological analysis and  EUV observations imaging the radiative signatures, we propose a complete 3D dynamics of the magnetic field configuration that describe the evolution of the flare.

We reconstruct the magnetic field of the event by using a NLFFF extrapolation and analyse its topology by computing the Q-factor. This allows us to identify most of the magnetic structure of the active region. First, we find a null-point, its associated separatrices, the fan and the spines, and the QSL-halos surrounding them. By confronting the separatrix footpoints and the ribbon location, we obtain an excellent match between the fan footpoints and the circular ribbon. While the outer spine footpoint is not perfectly co-spatial with the remote ribbon, the mismatch is only of 14 Mm which is a much smaller distance compared to any other studies so far. This almost perfect overlap between the ribbons and the separatrix footpoints implies that the reconstructed magnetic field is indeed very close to the real one. This reveals that our NLFFF extrapolation is able to accurately reproduce both an acceptable current distribution as well as the large scale magnetic field configuration.

In addition to the null-point topology embedded in its QSL-halo, we identify the presence of a flux rope and its associated HFT. 
Finding flux rope in eruptive events is quite common. However, investigating the origin of a confined flare, which are usually weak flares,  has not been an active research topic over the past few years. Flux ropes have been first identified as the trigger of a confined flare  by \cite{Wang_al15,Liu_al16}. In the present event, we confirm the role of a flux rope as the driver of confined flares, and for the first time in a null-point topology.  Such result suggests that flux ropes may be present at all scale and essential to any impulsive solar event, whether it is eruptive, confined or jet-like. Indeed, a flux rope carries current and indicates the non-potentiality of the system which should help to destabilise it.

The presence of an HFT below the flux rope has been highlighted several times in observations using the Q-factor computation \citep{Savcheva_al12, Zhao_al14}. Nonetheless, in our study, we show that prior to the main null point reconnection flare, reconnection occurs along the HFT. In addition to the co-spatiality of the HFT fooptrints and field lines with the EUV bright kernels and loops, we show that intense vertical electric currents are concentrated at the HFT footpoints. This indicates that HFT is loaded by electric currents which is a undeniable proof that magnetic reconnection occurred inside the HFT.

 As expected for reconnection at the null-point, a circular ribbon, an inner and a remote elongated ribbon are formed. We also observe the propagation of brightening along those 3 ribbons which is fully consistent with some slipping reconnection across the null-point QSL-halo \citep{Masson_al09b}. Moreover, using our topological analysis to interpret the EUV emission along the post-flare loops we show that the flare should mainly involve reconnection at the null-point between the flux rope in green and the overlying magnetic flux  in purple and red (see Figure~\ref{fig8}).
 Based on those results, we propose a scenario for the confined flare. In response to the photospheric flux injection, magnetic reconnection develops along the HFT associated to the flux rope. This should imply a growth of the flux rope as predicted by the standard 3D model of solar eruption \citep{Janvier_al13}. As a consequence, the flux rope rises, and it eventually reaches the null-point and the fan separatrix. Then the flux rope reconnects with the overlying flux above the null point. It leads to 1) the formation of the intense brightening on the North East of the circular ribbon (R1, \S~\ref{UV ribbons}) and 2) the post-reconnected loops undergoing slipping reconnection in the QSL-halo. In the meantime, reconnection at the null point occurs between the flux below the fan (materialised by the yellow field lines on Figure~\ref{fig4}) and the overlying field which leads to the formation of the circular ribbon itself.
 
 
In addition to the flaring phase, we also obtain informations on the post-flaring phase. In previous studies, the sets of post-flare loops were directly associated with the eruptive nature of the eruption. In \cite{Woods_al11} and \cite{Hock_al12}, the observed late EUV phase is co-temporal with the appearance of a large scale post-flare loops resulting from reconnection of the large closed loops initially confining the flux rope. This post-flare loops being very long, brightened several hours after the main flare phase. A different example of EUV late phase related to post-flare loops has been studied by \cite{Sun_al13}. In their event, they identified two post-flare loops, one resulting from the null-point reconnection acting as a breakout reconnection, i.e., removing the  flux above the flux rope but enclosed below the fan and the second as a consequence of the flux rope eruption itself, similarly to the two previous studies. 
The particularity of our event relies on its non-eruptive nature which is responsible for  the formation of 3 sets of post-flare loops of different length. Those sets of post-flare loops are temporally correlated with EUV late phases observed in the AIA light curves in cold and warm channels. The thermal evolution of each of the post-flare loops combined with the order of appearance in function of the length of those loops strongly indicate that a cooling process is at the origin of the 3 observed EUV late phases.

Contrary to eruptive flare \citep{Sun_al13}, here, the flux rope does not erupt but instead reconnects with the overlying field. 
In the present event, it is most likely that the reconnection processes most of the magnetic flux of the flux rope  and  prevents it to erupt. While we cannot confirm this statement, we can still discuss its implication with respect to the triggering mechanism of a flux rope eruption. Two competitive models have been proposed to explain the ejection of a flux rope : the torus instability \citep{TorokKliem07} and the breakout model \citep{Antiochos_al99}. The torus instability requires that the flux rope reaches an altitude with a rapid decrease of the magnetic field with height. The breakout model proposes that the overlying flux needs to be remove while the energy is building up in the flux rope itself in order to inverse the force balance of the system. In both scenarios the overlying magnetic field has to be small enough to allow the flux rope to erupt. In our studied event, the null point is confined deep in the corona and the amount of magnetic flux above the null-point might be too strong for the flux rope to get through. Instead it entirely reconnects with the overlying field failing to produce an eruption. This suggests that the flux ratio between the flux rope and the overlying flux may be important to determine the eruptive nature of a flux rope. \cite{Liu_al15} studied an events where the flux rope was not confined homogeneously. 
Section of the flux rope with a high ratio of the flux rope magnetic field on the confining field erupts while the flux rope section with a low ratio remain confined
The part of the flux rope with a strong field compared to the overlying field erupts while the part strongly confined by the strapping field did not. 

Circular ribbons and null-point topologies are also typically associated with solar coronal jets \citep{Moreno_al08,Wang_al12,Pariat_al09,MorenoGalsgaard13,Pariat_al15a}. Observational evidences accumulate for the presence of a flux rope in the pre-jet magnetic configuration \citep{Raouafi_al10,Sterling_al15}. Some of the  magnetic structures present in this confined event are typical of jet-like events. The magnetic flux rope reconnecting with the overlying field should transfer its twist to the outer magnetic field. This could lead to the generation of a propagating torsional Alfv\'en waves, that can be observed as a jet \citep{Pariat_al09,Pariat_al15a,ArchontisHood13}. Why in the present event no jet is observed and the dynamics of the event yields to a confined flare?

The answer is likely to be related to the geometry of the magnetic configuration. Jets usually develop when the surrounding field is uniform with respect to the typical scale of the closed source region \citep{Shimojo_al01}. In the present configuration, the length of the outer spine, L, is of the same order than the width, N, of the fan region.  In a recent MHD numerical study,  \citet{WyperDeVore16} show that jet formation depends on the ratio L/N. Clear jets preferentially developp for a large L/N ratio. In the present event, the L/N ratio is close to the unity which prevents the system to launch a propagating torsional Alfv\'en waves, i.e., no jets were observed. It is interesting to note that in the simulations of \citet{Edmondson_al09,Lynch_al14}, a non-linear propagating torsional Alf\'en wave appears immediately after the null-point switched from closed (with a low L/N ratio) to open (with a infinitely large L/N ratio) ambient magnetic field.  \citet{Wang_al12} present an observation in which  a confined flare configuration with a circular and an outer ribbon followed by a jet event are sequentially observed. They proposed that this transition from confined flare to jet results from the opening of the outer spine during the event. This is equivalent to the transition from a null point topology with a small ($\ll1$) ratio to a null-point with a large ($\gg 1$) L/N ratio.

 For small-scale non-eruptive flares such as the one studied here, the association of a flux rope confined in a null point topology may be relatively common. Magnetic null point appears when magnetic flux tube emerges in a pre-existing dipolar region, creating a quadrupolar configuration \citep[e.g.][]{Torok_al09}. In our case, the confinement of the flux rope by the null-point helps most likely to energize the system and, later, to release the flare energy impulsively. Indeed, by restraining the expansion of the flux rope, the null-point allowed to build up free energy in the system, which forced reconnection at the HFT and at the null-point. The null-point most likely plays a critical role in storing and releasing impulsively magnetic energy for flares or eruptions associated with a flux rope  \citep{Janvier_al16, Joshi_al15}. While this requires dedicated theoretical and numerical studies, it opens nonetheless a challenging issue concerning the generic role of a separatrix surface confining the eruptive structure (i.e., flux rope) on the trigger of solar eruption.

Overall, this study presents some evidences that a flux rope confined below a null point could be a relatively universal feature for impulsive events. Indeed, it can lead to large scale eruptive flares, solar jets and confined flares. This suggests that the geometry of the magnetic 
field surrounding the null-point configuration may be fundamental in determining the dynamics of the eruption. The relative size and strength of the surrounding 
magnetic field compared to the amount of flux inside the null point domain could dictate whether the erupting flux rope will develop as a full CME or will be 
destroyed though magnetic reconnection. In the later case, the ratio between the fan width and the outer spine length would define whether the event becomes a jet or a confined flare.

\section{Acknowledgements} 
The authors are very grateful to the HMI team and in particular to Xudong Sun for providing the processed HMI data used for the extrapolation and the soft to remap the data to the CEA projection.
GV acknowledges the support of the Leverhulme Trust Research Project Grant 2014-051. CL, ND and HW are supported by  NASA under grants NNX13AF76G, NNX13AG13G, NNX14AC12G, and NNX16AF72G,  and by NSF under grants AGS 1250818, 1348513 and
1408703. HR acknowledges support from the STFC consolidated grant ST/L000741/1
and from a SUPA Advanced Fellowship.
Calculations were done on the quadri-core bi-Xeon computers of the Cluster of the Division Informatique de l'Observatoire de Paris (DIO).
The authors thanks the ESA JHelioviewer team for the JHelioviewer tool.

\bibliographystyle{aa}
\bibliography{../../phd_biblio}  
\IfFileExists{\jobname.bbl}{} {\typeout{}
\typeout{***************************************************************}\typeout{***************************************************************}\typeout{** Please run "bibtex \jobname" to obtain the bibliography} 
\typeout{** and re-run "latex \jobname" twice to fix references} 
\typeout{***************************************************************}\typeout{***************************************************************}\typeout{}}

\end{document}